%% file: main.tex
\documentclass[sigconf]{acmart}

\AtBeginDocument{%
  \providecommand\BibTeX{{%
    \normalfont B\kern-0.5em{\scshape i\kern-0.25em b}\kern-0.8em\TeX}}}

\copyrightyear{2023} 
\acmYear{2023} 
\setcopyright{rightsretained} 
\acmConference[FAccT '23]{2023 ACM Conference on Fairness, Accountability, and Transparency}{June 12--15, 2023}{Chicago, IL, USA}
\acmBooktitle{2023 ACM Conference on Fairness, Accountability, and Transparency (FAccT '23), June 12--15, 2023, Chicago, IL, USA}
\acmDOI{10.1145/3593013.3594073}
\acmISBN{979-8-4007-0192-4/23/06}

\usepackage{graphicx,verbatim,tabularx,setspace,xspace}
\usepackage{xcolor}
\usepackage{array}
\usepackage{boldline}
\usepackage{subcaption}
\usepackage{multirow}
\usepackage{pdflscape}
\usepackage{tabularx,ragged2e}
\usepackage{pifont}%
\usepackage{rotating}
\usepackage[utf8]{inputenc}
\usepackage{geometry}
\usepackage{ragged2e}
\usepackage{booktabs, tabularx}
\usepackage{enumitem}
\usepackage{etoolbox}
\usepackage{makecell}

\usepackage{pgfplots}
\pgfplotsset{compat=newest}
\usetikzlibrary{patterns}

\usepackage{url}
\newcolumntype{L}[1]{>{\raggedright\arraybackslash}p{#1}}

\newcolumntype{C}[1]{>{\centering\arraybackslash}p{#1}}

\newcolumntype{R}[1]{>{\raggedleft\arraybackslash}p{#1}}

 	 \def\changejs{}
    
    \def\changejc{}
    
\def\cut#1{}

\usepackage{etoolbox} %
\newtoggle{comments} %
\iftoggle{comments}{
    \newcommand{\notejs}[1]
    	{{\color{orange}[{\bf Jat:} #1]}}
    \newcommand{\notemv}[1]
    	{{\color{purple}[{\bf Michael:} #1]}}
	\newcommand{\notejc}[1]
    	{{\color{violet}[{\bf Jennifer:} #1]}}
    
    \newcommand{\todo}[1]
    	{{\color{red}[{\bf Todo:} #1]}}	
    	
    \newcommand{\cut}[1]
    	{{\color{red}\sout{#1}}}
       
}{
    \newcommand{\notejs}[1]{}
    \newcommand{\notemv}[1]{}
    \newcommand{\notejc}[1]{}
    \newcommand{\todo}[1]{}
}

\begin{document}

\title{Understanding accountability in algorithmic supply chains}

\author{Jennifer Cobbe}
\orcid{0000-0001-8912-4760}
\affiliation{%
  \institution{University of Cambridge}
  \country{United Kingdom}
}

\author{Michael Veale}
\orcid{0000-0002-2342-8785}
\affiliation{%
  \institution{University College London}
  \country{United Kingdom}}

\author{Jatinder Singh}
\orcid{0000-0002-5102-6564}
\affiliation{%
  \institution{University of Cambridge}
  \country{United Kingdom}}

\renewcommand{\shortauthors}{Cobbe, Veale and Singh}

\begin{abstract}
Academic and policy proposals on algorithmic accountability often seek to understand algorithmic systems in their socio-technical context, recognising that they are %
produced by `many hands'. Increasingly, however, algorithmic systems are also produced, deployed, and used within a supply chain comprising multiple actors tied together by %
flows of data between them. In such cases, it is the working together of an \textit{algorithmic supply chain} %
of different actors who contribute to the production, deployment, use, and functionality that drives systems and produces particular outcomes. %
We argue that algorithmic accountability discussions must consider supply chains and the difficult implications they raise for the governance and accountability of algorithmic systems. %
In doing so, we explore algorithmic supply chains, locating them in their broader technical and political economic context and identifying some key features that should be understood in future work on algorithmic governance and accountability %
{(particularly regarding general purpose AI services).}
To highlight ways forward and areas warranting  attention, we further discuss some implications raised by supply chains: {challenges for allocating accountability stemming from \textit{distributed responsibility} for systems between actors, limited visibility due to the \textit{accountability horizon}}, service models of use and liability, and cross-border supply chains and regulatory arbitrage.

\end{abstract}



\maketitle

\input{s1-intro.tex}

\input{s2-algo-accountability.tex}
\input{s3-supplychains.tex}

\input{s4-implications.tex}

\input{s5-conclusion.tex}

\begin{acks}
     JC and JS are members of the Compliant \& Accountable Systems Group, which acknowledges the financial support of UK Research \& Innovation (EP/P024394/1, EP/R033501/1, ES/T006315/1), The Alan Turing Institute, and Microsoft (via the Microsoft Cloud Computing Research Centre). MV is supported by the Fondation Botnar.

\end{acks}

\balance
\bibliographystyle{ACM-Reference-Format}

\end{document}

%% file: s1-intro.tex
\section{Introduction}
\label{sec:introduction}

The ‘many hands’ problem holds that accountability is difficult where many people have together contributed to activity or outcome, as it may be impossible to allocate responsibility to any one of them~\cite{bovens1998,thompson1980,nissenbaum1996}. Writing in 1996, Nissenbaum argued that computer systems raise a particular form of this problem -- they are usually %
produced by groups or organisations rather than individuals, and may include components %
developed by others~\cite{nissenbaum1996}. Addressing the `many hands' problem is increasingly a concern in the algorithmic accountability literature (implicitly or explicitly), with proposals in recent years for accountability for algorithmic systems to operate at an organisational level~\cite{cobbe2021reviewable,krollaa,raji2020closing,williams2021,madaio2020co,gebru2018datasheets,mitchell2019model,singh2016partofaprocess,kroll2021traceability}.  %

Yet today's computer systems---including AI technologies---are increasingly modular, %
\changejc{dependent on} cloud-based technologies, and interconnected. The `agile turn' of %
recent decades %
\changejc{transformed} software development, distribution, and infrastructure, directly influencing how businesses are organised and computing resources are distributed~\cite{gurses2017}. %
\textit{Agile development} means %
software {(including `AI' models)} is now produced in short development cycles with continuous testing %
\changejc{and iterative revision} after deployment~\cite{gurses2017}. %
\changejc{Computing resources are} now generally modularised and distributed \textit{as a service}, with a client-server model in which the server performs the computation%
~\cite{gurses2017,millard2021}. The challenges of scaling services and %
increasingly portable %
client devices %
\changejc{drove} advances in data centres with flexible resources and software becoming increasingly \textit{cloud-based}~\cite{gurses2017,millard2021}.
 Consequently, software development now often involves, to various degrees, integrating pre-built modular components provided \textit{as services} and controlled by others into a complete product: not simply a system, but %
 a \textit{system-of-systems}~\cite{singh2019decprov}.%

As a result, digital technologies across society and the economy are increasingly organised around \textit{data-driven supply chains} involving several interconnected actors and their systems. In these supply chains, data flows %
between actors, linking systems designed, developed, owned, and controlled by different people and organisations%
~\cite{norval2020reviewableIoT}: a sensor system (controlled by one actor) might connect to an analytics system (controlled by another) %
which itself outputs into a decision-making system (controlled by a third).  %
This is often so even for seemingly simple applications; for example, a home thermostat can be driven by data from a national weather %
service, which is itself fed data from thermometers owned and operated by different actors. In such supply chains, the \textit{working together} of %
services and systems {controlled by different actors} produces particular outcomes---hardware capabilities, software functionalities, the workings of commercial and industrial processes, ‘AI’ decisions and outputs, and more. Supply chains are \textit{data-driven} in that the flow of data between actors links them together, allowing a system controlled by one actor to interact with those controlled by 
others and together produce some functionality~\cite{singh2019decprov,cobbe2019whatlies}.
In the context of AI and algorithmic systems, \textbf{algorithmic supply chains} are those where several actors contribute towards the production, deployment, use, and functionality of AI technologies. In these supply chains, AI ‘as a service’ providers often play key roles~\cite{cobbe2021artificial}. 

{By} reconfiguring software %
\changejc{production} and distribution, the agile turn also had significant political economic \changejc{and other} ramifications 
~\cite{gurses2017,cobbe2019whatlies}. In bringing services together to produce functionality through supply chains, developers now delegate control over much of the underlying technologies to 
\changejc{others}, complicating the governance of those technologies and the products they are part of. It is no longer necessarily %
\changejc{true} that computer systems are %
produced by a group of developers or an organisation, or by {a vendor} simply integrating various %
standalone components into one product (%
\changejc{itself raising} the `many hands' problem~\cite{nissenbaum1996}). Instead, %
they %
often now %
involve a group of organisations arranged together in a data-driven supply chain, %
{\textit{each retaining control over component systems they 
provide as services to others}}. Moreover, certain key actors in %
supply chains---in particular, major cloud providers who often control underlying technologies---provide many %
services to millions of customers, %
holding important positions across supply chains in many sectors~\cite{cobbe2019whatlies,cobbe2021artificial}. The agile turn %
has \changejc{thus} reorganised many areas of social and economic life -- now reconstituted around %
\changejc{data-driven} supply chains with a few systemically important actors providing the core infrastructure that underpins contemporary society. 

Algorithmic supply chains bring %
significant implications for governance and accountability frameworks and mechanisms relevant to algorithmic systems%
. Allocating %
accountability across supply chain actors for producing, deploying, and using algorithmic systems is relevant to general academic, policy, and regulatory discussions %
around algorithmic accountability, %
and to more specific legislative efforts around regulation of AI. %
Here we argue that \textit{governance of and accountability for algorithmic systems} as deployed and used in the real world %
\changejc{must} \textit{operate across the supply chains} %
\changejc{which} will increasingly underpin, drive, and produce the\changejc{ir} outputs and effects.%

Much of the policy and academic literature, however, is grounded in an organisation-focused understanding of digital technologies. Even %
recent %
work which seeks to address the `many hands' problem through a relatively broad view of accounting for 
algorithmic systems is typically focused on making specific stages of system lifecycle more transparent %
~\cite{gebru2018datasheets,mitchell2019model,arnold2019factsheets,crisan2022interactivemc,pushkarna2019datacards} or framed around the perspective of a single organisation~\cite{cobbe2021reviewable,
wieringa2020account,raji2020closing,williams2021}. This attention now paid to organisations' accountability %
for their algorithmic systems was long overdue, but the focus on organisational accountability has largely obscured the dynamics of %
algorithmic supply chains. We therefore still lack ways to conceive of these chains, to bring them within legal, regulatory, and governance mechanisms, and to appropriately distribute responsibility and accountability. %

This paper contributes to understanding these challenges. First (\S\ref{sec:algosystemaccountability}), we discuss recent trends in algorithmic accountability and identify %
limitations regarding supply chains. Next (\S\ref{sec:supplychains}), we describe \changejc{and contextualise} AI services and algorithmic supply chains %
and identify %
key features of how they are structured and operate%
. Then (\S\ref{sec:supplychainaccountability}) we discuss important implications of these features for accountability%
: the distributed nature of responsibility in supply chains (\S\ref{sec:distributedresponsibility}); the limited understanding individual actors may have of the broader chain {due to the {`}accountability horizon{'}} (\S\ref{sec:accountabilityhorizon}); %
and providers' efforts to %
structure supply chains to maximise control and commercial advantage while minimising %
legal risk and accountability (\S\ref{sec:benefit}).

In all, we argue, algorithmic accountability work must urgently address the technological, legal, and political economic dynamics of algorithmic supply chains. We do not offer concrete %
\changejc{technical or organisational proposals} to improve accountability in these supply chains, but instead hope to produce a shift in focus for algorithmic accountability as a field and indicate new research directions

%% file: s2-algo-accountability.tex
\section{Accountability in algorithmic systems}
\label{sec:algosystemaccountability}

Significant academic and policy work has sought various forms of accountability %
for algorithmic systems~\cite{williams2021}. Accountability %
\changejc{is often} seen either as a \textit{mechanism} \changejc{(particularly in Europe and non-US Anglophone countries)} or a \textit{virtue} \changejc{(particularly in the US)}~\cite{bovens2010}. As a mechanism%
, it is an institutional arrangement whereby an \textit{actor} provides %
accounts to a \textit{forum}, who deliberates on %
those accounts and may impose consequences~\cite{bovens2006}. A developer might provide information to a regulator about their system, for example, with the regulator then issuing a penalty or requiring design changes. Some algorithmic accountability literature explicitly views accountability as a mechanism for holding actors to account for \changejc{their} systems%
~\cite{cobbe2021reviewable,wieringa2020account,adalovelace2020,krollAccountabilityComputerSystems2020}. By contrast, accountability as a virtue %
is a normative concept, a set of standards for evaluating behaviour---often tied to being transparent, responsible, and responsive---with ‘being accountable’ seen as a positive quality of particular actors~\cite{bovens2010}. Some (predominantly technical) work has %
thus sought to improve the accountability of certain technologies by imbuing them with such positive qualities. Yet applying accountability to algorithmic systems in this way---rather than to the organisations responsible for them---often equates accountability with technical functionality (for example, building `Accountable AI') rather with human virtues which are not reducible to technically tractable concepts~\cite{krollAccountabilityComputerSystems2020}.

We treat accountability as a mechanism, whereby  actors are held accountable for technologies they are responsible for. However, accountability for digital technologies is %
\changejc{often} challenging. The `many hands' problem---%
\changejc{that often no one person is responsible for }
outcomes which multiple people helped produce%
---has long been recognised: computer systems are rarely produced by an individual who can be held %
\changejc{accountable}, but by teams and organisations with many people contributing~\cite{nissenbaum1996}. Moreover, modular software development---where software developed by one organisation uses a library developed by another, for example---further complicates things~\cite{nissenbaum1996}. Much software is too complex, relying on too many components, for any one person to %
account for all of its workings.

In the context of algorithmic accountability, specifically, a key conceptual shift has been in understanding these systems not as ‘algorithms’ but as %
\textit{algorithmic systems}: “intricate, dynamic arrangements of people and code”~\cite{seaver2013}. This recognises that ‘algorithms’ are produced and work within %
human contexts and in practice cannot be understood separately from them. Simultaneously,
explanations of (ML) model workings are increasingly recognised as insufficient to account for algorithmic systems~\cite{edwards2017slave,cobbe2021reviewable,wieringa2020account}. Much research has therefore gradually moved away from seeking transparency or explanations of models (though this remains an important area of work) to understanding algorithmic systems more broadly as socio-technical phenomena. Much of this reflects---implicitly or explicitly---an understanding that algorithmic systems are often the result of `many hands': produced by and deployed and used within teams and organisations. To %
account for an algorithmic system, one needs to %
account for the collective efforts of the organisational processes involved in producing, deploying, and using it.

The term ‘algorithmic system’ is now widely used in algorithmic accountability%
, with academic and policy literature commonly suggesting ways to improve accountability for their organisational aspects. Some proposals seek lower-level mechanisms to document the choices and decisions made by people in developing, deploying, or using a system, such as %
for datasheets~\cite{gebru2018datasheets} or data cards~\cite{pushkarna2022datacards} to describe datasets, or model cards~\cite{mitchell2019model,crisan2022interactivemc} and factsheets~\cite{arnold2019factsheets} to describe model specification and capabilities. Such proposals often recognise accountability as a positive quality (i.e. a virtue) and seek %
improved transparency of algorithmic production and deployment processes. %
\changejc{Higher-level proposals seek} to integrate lower-level mechanisms and provide ways of understanding %
holistically the \textit{process} of producing, deploying, and using %
algorithmic systems, such as %
for auditability \cite{williams2021}, reviewability~\cite{cobbe2021reviewable}, contestability~\cite{kaminski2021}, traceability~\cite{kroll2021traceability}, and others~\cite{wieringa2020account}. %
\changejc{These} have mainly reflected accountability as a mechanism, and sought ways to support institutional mechanisms and accountability relationships between actors and forums. Though coming from different perspectives, these various lower- %
and higher-level mechanisms %
all essentially %
\changejc{recognise} that algorithmic accountability---either as a virtue or a mechanism---must reflect the `many hands' nature of AI technologies.

More recently, a `second wave' of algorithmic accountability research has sought to address more structural concerns around the development, deployment, and effects of algorithmic systems~\cite{pasquale2019}. This work moves from creating better methods %
to scrutinise systems \textit{in situ} to considering \textit{whether} such systems should be built at all, how, %
why, and who gets to govern them. This %
echoes longer-standing critical work in fields such as surveillance studies, which has considered the structural impacts of technologies of sorting and profiling on societies, and in which arguments exist against using these technologies altogether \cite{gandyEngagingRationalDiscrimination2010,harcourtPrediction,lyonSurveillanceSocialSorting2003}. While literature in these fields considers issues such as the cumulative effects of systems on individuals and communities \cite{gandyEngagingRationalDiscrimination2010}, they typically consider systems themselves through an organisation-centric lens -- equating %
\changejc{particular functionality} (credit scoring, criminal profiling, airport screening, targeting advertising), with either the actor authorising the action (bank, police department, interior ministry, online platform), or %
a particular technology provider or contractor.

Yet following the agile turn, an organisation-centric view is {a less meaningful frame for analysis.} %
Consequential algorithmic systems are commonly produced, deployed, used, and have effects through and within supply chains. It is therefore no longer the case that 
software is generally developed by particular teams or organisations (who may have integrated components developed by others into their finished product). Instead, as we argue, functionality results from the \textit{working together of multiple actors} across various stages of production, deployment, and use of AI technologies (connected by data flows across organisational, legal, technical, visibility boundaries). %
This does not mean that a particular single organisation will never be appropriate to hold to account, but that identifying the actors and processes that led to the functionality of any particular algorithmic system becomes significantly less straightforward. %

We next (\S\ref{sec:supplychains}) explore key features of algorithmic supply chains, followed by their challenges and implications for the mechanism of algorithmic accountability (\S\ref{sec:supplychainaccountability}).

%% file: s3-supplychains.tex
\section{AI services and supply chains}
\label{sec:supplychains}

Significant barriers to entry limit the number of organisations that can produce bespoke state-of-the-art AI technologies in-house, either for their own use or to bring to market~\cite{cobbe2021artificial}. Developing, maintaining, and renewing advanced AI technologies typically requires large and relevant quantities of data, potentially from multiple sources and labelled or moderated, relating to many use-cases, contexts, and subjects. Cutting-edge model development requires scarce expertise in model training, testing and deployment, all with significant storage, compute, and networking needs.

Companies with these capabilities now offer commercial access to cloud-based AI technologies `as a service’ (AIaaS)~\cite{cobbe2021artificial,lewicki2023aiaasfairness}. Major companies including Amazon~\cite{AmazonAIServices}, Microsoft~\cite{MicrosoftCognitive}, Google (Alphabet)~\cite{GoogleCloudAI}, and IBM~\cite{IBMWatsonAI} offer networked access to various state-of-the-art AI capabilities, including both model-building services and pre-built (and `general purpose') %
models in areas such as language, speech, vision, and analytics (see~\cite{lewicki2023aiaasfairness}){, or {generative models for producing} %
text, images, audio, or video. }%
Some companies offer specific services to customers, such as facial recognition~\cite{clearviewAI}, hiring~\cite{HireVue, pymetrics}, or medical diagnostics~\cite{Lunit75:online, Infermedica}. {And some operate as platforms for all the above, looking to connect developers, clients and infrastructure providers, among others, in a multi-sided market -- Amazon and Microsoft, for example, offer access to models from other providers alongside their own~\cite{awsintermed,msopenai}, whereas other platforms are {primarily} an intermediary (such as HuggingFace~\cite{HuggingFace})}. %
 AI services can be integrated into apps and Web services, analytics systems, business and industrial processes, workflows, and with IoT devices with real-world physical effects (collectively: `\textit{applications}'). Low marginal cost and effort means this will likely become the primary way that organisations integrate AI capabilities~\cite{cobbe2021artificial}.

\begin{figure*}[ht]
  \centering
  \includegraphics[width=.85\linewidth]{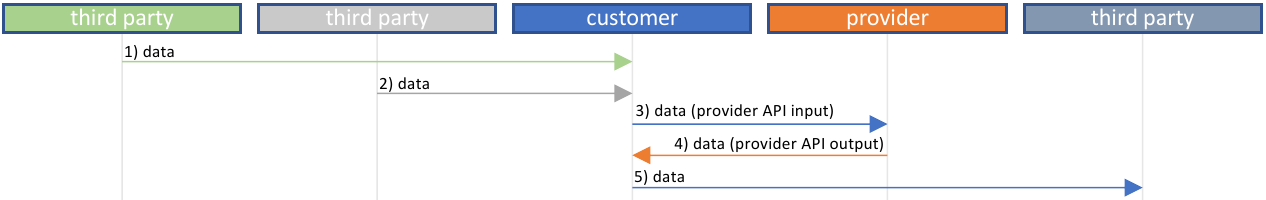}
  \caption{Sequence diagram of a simplified data-driven supply chain with an AI service. The customer sends input data to provider's API, who performs some %
  computation%
  , before returning the results to the customer. Some of the broader supply-chain is illustrated, where the customer has previously received data from other third parties, and later, sends some data to another.}
  \label{fig:thirdparties}
  \vspace*{-11pt}
\end{figure*}

AI services take various forms~\cite{lewicki2023aiaasfairness}. Here we focus on services offering access to pre-built `general purpose' models and to customised models tailored using tools offered by providers. In these, providers take major roles in the technology's production and distribution, developing and hosting systems on their (owned or managed) infrastructure. Services are typically accessed through application programming interfaces (`APIs') {controlled by providers}, which allow the underlying system's capabilities to be integrated into applications {by customers} {(Fig. \ref{fig:thirdparties})}. {This client-server model thus allows providers' algorithmic systems to run on their infrastructure, under their control, even while they are deployed by customers in applications across many contexts and use-cases.} There are typically few (if any) checks on customers' identities or intentions, services use standard-form contracts (at least for smaller clients), and customers are billed on the 
API calls made~\cite{cobbe2021artificial}.

An application's supply chain may involve several AI and non-AI services%
, potentially from multiple providers. {Indeed, an AI service may be only one part of a broader, more complex %
chain for a given application, %
which may integrate multiple AI and other services}. Actors in these chains are broadly `upstream' or `downstream' from the perspective of others -- though this distinction can blur where actors take multiple positions in a chain. AI services themselves have supply chains, such as for dataset production activities like data gathering and labelling~\cite{matusCertificationSystemsMachine2022} (see \S\ref{sec:arbitrage})\changejc{,  typically involving data from customers' application deployments~\cite{cobbe2021artificial} and from providers' own user-facing platforms and services. Customers therefore appear in supply chains for deployment \textit{and} production of some AI services, while end-users of customers' applications and providers' services are themselves drawn into production}. Firms using AI services may themselves provide services to their own customers~\cite{millard2021}, such as proctoring software sold to universities repackaging Amazon's facial recognition service~\cite{amazonCaseStudyWeShine2021}, or copywriting software repackaging OpenAI's text generation model (GPT-3 %
~\cite{ArtificialIntelligencePermeating2022}). %

We %
identify several key features of algorithmic supply chains:
\begin{itemize}
\item production, deployment, and use are split between several interdependent actors; 
\item supply chain actors and data flows perpetually change; 
\item major providers' operations are increasingly integrated across markets and between production and distribution; and 
\item supply chains are increasingly consolidating around systemically important providers. 
\end{itemize} We draw out their implications for accountability in \S\ref{sec:supplychainaccountability}.

\subsection{Supply chains split production, deployment, and use between %
interdependent actors}
\label{sec:interdependence}

\begin{figure*}[htb]
  \centering
  \includegraphics[width=.6\linewidth]{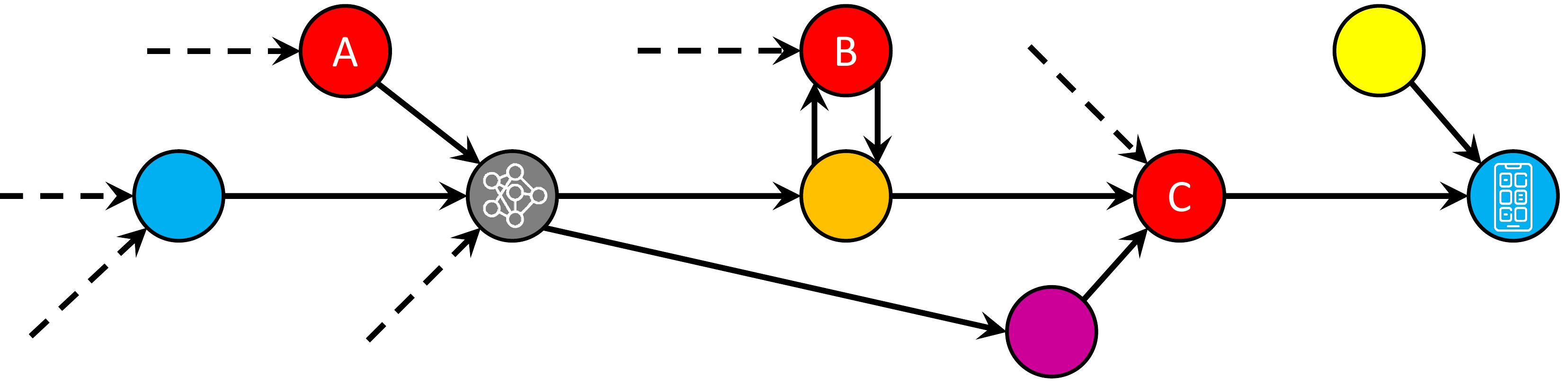}
  \caption{A representative AI supply chain. The application developer (blue) initiates a series of data flows by sending input data to an AI service provider (grey). One AI service provider (red) appears at multiple key points in the supply chain – providing infrastructure (A) for an AI service offered by (grey); providing an AI service (B) to another cloud service provider (orange); and providing technical infrastructure (C) for application deployment. %
  }
  \label{fig:multipoint}
\end{figure*}

In algorithmic supply chains, different aspects of production, deployment, and use of AI technologies are split between multiple actors tied together by data flows. 
{As a result, the activities of the various actors in supply chains each depend on {the actions %
of } %
by  others}.  %
This may involve various {interacting} AI and non-AI technologies, such as cloud services, servers, %
data centres, data sources, and content delivery networks, controlled by different actors. %
The \textit{working together} of the various actors {who control these technologies}---each doing something that enables, supports, or facilitates the actions of others---produces a particular outcome (see Fig. \ref{fig:multipoint}). %
Each actor in a supply chain may not be aware of the others, nor have consciously decided to work together towards that outcome -- indeed, they may have limited understanding of actors even one or two steps removed (see \S\ref{sec:accountabilityhorizon}). However, each depends on something done by others, and their role in a supply chain is contingent on the activities of actors both up-- and downstream of them.

Actors in algorithmic supply chains are thus \textit{interdependent}, each doing something to fulfil the needs of others (such as processing a particular data input and returning an output, or providing infrastructure to support application deployment). %
The interdependence of the various actors responsible for developing, deploying, and operating algorithmic systems in supply chains means  they are not individual, independent actors \textit{as such}. Instead, these actors, their relations, and their role in the workings and effects of AI technologies can only be understood \textit{in the context of that supply chain}. Studying an actor and their systems in isolation from supply chain contexts is akin to studying an algorithmic model in isolation from its broader organisational context (the limitations of which are increasingly recognised (see \S\ref{sec:algosystemaccountability})). The dynamics of interdependence in algorithmic supply chains---how they are structured, the relative importance of actors, and how problems spread---are therefore key considerations for algorithmic accountability, as we now explore.

\subsubsection{Supply chain interdependencies are structured by technological, legal, and political economic factors}
\label{sec:structuringinterdependence}

Interdependence between actors gives algorithmic supply chains their structure and functionality. Certain actors---typically (AI and non-AI) cloud service providers---have leveraged AI technologies they own, production processes they control, and cheaply-accessed networking technologies to pursue particular interdependencies with others and strategically position themselves in markets and supply chains of many kinds. Technologies afford certain capabilities to those who use or control them~\cite{gibson1979,norman1988,davis2020}. They can therefore also afford the ability to do things that fulfil the needs of others. Because, as we discuss in \S\ref{sec:interdependence}, people doing things for each other produces interdependence between them~\cite{elias1984}, different technologies can afford different kinds of interdependencies. Networking and data processing technologies, for example, allow the stages of production and deployment of AI technologies to be distributed geographically. They can therefore be done by different people, each of whom does something for the others, producing interdependence between them.

However, technologies and their affordances cannot \textit{determine} interdependencies or the structure of supply chains. Affordances are not objective properties of technologies, but depend on context and perspective~\cite{gibson1979,norman1988,davis2020,diver2018,verbeek2005}. How providers can strategically position themselves is thus shaped both by their technologies' affordances \textit{and} by social, legal, and political economic factors %
which also influence how %
actors relate to each other, what they do for each other, and the interdependencies that arise. Accordingly, to position themselves in supply chains and markets, providers have also leveraged political economic factors such as economies of scale and
favourable legal frameworks such as intellectual property, intermediary liability, and data protection~\cite{cohen2019,cobbe2021artificial,cobbeDPDW}. Political economic and legal factors---not just technological---are thus important in producing and structuring algorithmic supply chains.

\subsubsection{Some actors are core players in supply chains}
\label{sec:coreandperiphery}

Supply chain actors are %
{generally}  not equal in their interdependence with each other, and some may do things that others particularly depend on. Their services, for example, may be relied upon by multiple others, {as is often the provider's aim in offering general purpose services}. Providers may depend on each customer only a little, while customers may depend on the provider for business-critical application functionality. Supply chain interdependencies are thus often \textit{asymmetric}, with certain actors---typically including at least those responsible for production of AI technologies---performing core functions for others, while others are more peripheral. Various contextual signs might indicate that an actor is core in a particular chain. For example, they may be the application developer who calls the supply chain into existence. They may perform some function (such as providing an AI service) which is crucial to application functionality. They may provide a key step between actors (such as offering access to another provider's technology through their API) upon which subsequent steps of the chain depend. Or they may appear at multiple different points in the chain providing cloud-based technical architecture on which functions performed by other actors rely. Some actors may also be more interchangeable and replaceable than others -- the barrier to entry for applying a specific off-the-shelf API is typically substantially lower than to generate the underlying technology to begin with.

These asymmetries of interdependence produce asymmetries in power~\cite{elias1984}: where one actor depends more on things done by another than that other depends on them, the balance of power between them favours the second actor~\cite{elias1984}. An application developer, for example, who uses a major provider's AI service, will depend more on that provider than the provider---who has many customers---will depend on that one developer. %
Power balances in algorithmic supply chains thus arise relationally yet asymmetrically and %
change over time as the relations and interdependencies between actors evolve~\cite{elias1984}. These power balances are not determined by actors' relations to the \textit{technologies} involved, but through their relation to and interdependence with \textit{each other}. As we note (\S\ref{sec:structuringinterdependence}), this is subject to many potential influences, of which factors like control of production processes and APIs are just some. But, by leveraging their technologies alongside legal, social, and political economic factors, providers can hold significantly asymmetrical positions %
as core actors in many supply chains, with power balances between them and others heavily in their favour.

\subsubsection{Interdependence helps problems propagate}
\label{sec:propagation}

Supply chain interdependencies mean %
problems with one actor’s technologies can propagate through other actors' systems. Where an AI service is biased in some way (such as facial recognition performing poorly on particular demographics~\cite{buolamwini2018gender}), that bias will be inherited by %
applications relying on that service~\cite{lewicki2023aiaasfairness}. Such a cascade's effects may be complex and unpredictable given the dynamic and largely undocumented %
set of actors and interdependencies found in many chains. %
\changejc{Statistical guarantees may not hold} when systems are composed together, and it is not straightforward to evaluate a whole system when each individual component may have been evaluated under different threat models (or other critiera)~\cite{kostovaPrivacyEngineeringMeets2020a, lewicki2023aiaasfairness}. %
Unless identified by the provider, actors `downstream’ from them may be unaware of a problem until they notice some unexpected behaviour. %
Even then, because they have delegated 
key aspects of production (and possibly deployment) of the technologies their application relies on to other actors (such as AI service providers), customers may struggle to understand where in a supply chain the problem has arisen, why, and what they can do to mitigate it. 

\subsection{Supply chains are transient and dynamic with unstable interdependencies}%
\label{sec:process}

Agile development combined with %
services-based distribution models %
has produced algorithmic supply chains which operate as dynamic \textit{processes} of data flow between a changing number and arrangement of actors. %
Just as critical engagement with algorithmic systems must recognise that they change over time~\cite{seaver2013}, so do algorithmic supply chains. %
 Indeed, because \changejc{data flow ties} %
 actors %
 together%
 , a chain may differ each time it is instantiated. At various points there may be multiple possible directions for data to flow between actors depending on the outcomes of analyses performed on it. 
 A face detected in a video stream using one service, for example, might trigger a flow to a separate facial recognition service to identify the person (%
 \changejc{with} its own supply chain %
 and associated data flows) and back again. %
 This might trigger a flow to a third system to record and alert of the presence of %
 a particular individual. %
Supply chains can thus be dynamically instantiated, and their structure may vary depending on the input data and the outputs of component systems. A supply chain's structure---the actors involved, what they %
do for each other, the interdependencies and power balances between them---may therefore only be fully apparent once the functionality or outcome has been been produced. %

However, technical, legal, or economic relations between actors do often persist across multiple instances of a particular supply chain. An application developed to use a particular provider's service will typically use that service repeatedly, even if the path of data flow between actors differs between instances. As such, while supply chains may change overall, %
bilateral relations between particular actors may remain relatively consistent. However, the nature of their relationship---the services provided and used, for example---may still change over time. An application developer may introduce new features, for instance, which use additional services offered by the particular provider. They may deprecate features such that particular services are no longer needed. They might employ additional support services for rapid growth in application resource requirements (for example, where an application `goes viral'). The provider may change their terms of service (altering the legal relationship between them) or withdraw particular services (resulting in changes in the developer's application). %
These are just some of the ways that relationships between actors may change.

\subsection{Some actors are integrated across markets and between production and distribution}
\label{sec:integration}

Some providers of AI and other cloud services commonly found in algorithmic supply chains have reached high levels of integration; both horizontally (across markets and sectors), and vertically (across production and distribution processes). This has implications for their positioning and role in algorithmic supply chains.

\subsubsection{Horizontal integration}
\label{sec:horizontal}

Horizontally integrated companies %
operate across markets and sectors. The most prominent cloud providers (Amazon, Microsoft, Google, Alibaba, IBM) offer {various} services %
across many related and adjacent markets and may appear repeatedly in a %
supply chain. Some %
such services are AI-related; others are infrastructure for applications (storage, database, content delivery, %
credential management, and so on); still others are user-facing, from business and consumer web-based services (such as maps or photo backup) to software packages for customers and their users (such as Microsoft 365). This allows a single provider to %
\changejc{re-purpose their AI and other technologies across} a range of services%
\changejc{, both infrastructural and user-facing.} 
It is %
common for providers to
purchase potential competitors and new market entrants, either to obtain intellectual property%
, to expand their services across markets, or to stifle emerging competition in existing markets. Providers can also simplify %
\changejc{how} existing customers %
bring AI services within their applications by providing tools to facilitate integrating them with their other services. Providers may financially incentivise customers to use several of their services instead of those of competitors.

\subsubsection{Vertical integration}
\label{sec:vertical}

Vertically integrated companies control multiple stages of production and distribution. Several major AI providers{---primarily Amazon, Microsoft, and Google---}own key infrastructure for %
\changejc{producing their systems and distributing them as} services across markets: data centres and servers; %
content delivery networks; APIs and customer-facing interfaces; and %
network infrastructure. %
Vertical integration offers %
bespoke technical infrastructure %
specific to %
these providers' needs which they can use for many services across markets to exploit economies of scale. High resolution media (requiring significant resources), for example, thus encourages vertical integration, as does state-of-the-art AI production (requiring more data, bigger and more complex models, intensive compute, and sophisticated training and testing processes). %
\changejc{Vertically integrated providers can link deployment of systems by customers to their production processes, testing and further refining} those AI technologies using customers’ input data, applications, and real-world use cases~\cite{cobbe2021artificial}. This allows providers to reduce the resources needed to improve models, while offsetting some research and development costs by bringing it into a process %
paid for by customers~\cite{cobbe2021artificial}. They can thus lower the net cost of developing more accurate and more generalisable systems~\cite{cobbe2021artificial}.

However, vertical integration has limits. AI providers might not operate %
in-house data cleaning and labelling processes, for instance (key parts of training, testing, and updating models). %
The business benefits to providers of bringing these processes ‘in house’ are potentially outweighed by the commercial advantages of extending supply chains across borders to exploit differences in laws  (\S\ref{sec:arbitrage}). %
Aspects of AI production are often \changejc{instead} %
outsourced to low-paid and insecure workers in the Global South~\cite{time2023,forbes2019} (%
\changejc{through data cleaning and labelling services} offered by companies like Sama AI~\cite{samaAI}, or through %
Mechanical Turk~\cite{mturk}). Moreover, some major providers
now offer access through their services to generative (foundation) models produced and controlled by others, marking a shift towards \textit{less} integration in some emerging product sectors.

\subsubsection{Providers all the way down}
\label{sec:othersuppliers}
Though some prominent AI providers are both horizontally and vertically integrated, most %
are not. Instead, they tend to specialise in a few closely-related services, such as algorithmic recruitment, processing legal documents, certain medical processes~\cite{lewicki2023aiaasfairness}, and even `algorithmic governance' and `ethical AI' (see~\cite{ethicaldb}), %
\changejc{without operating} across traditional cloud service markets. These {specialist}
providers typically %
‘rent’ technological %
infrastructure from a larger provider \changejc{rather than operating their own}
(OpenAI, for example, exclusively uses Microsoft’s Azure cloud services~\cite{msopenai}). %
{This reflects the fact that %
developing advanced AI technologies and operating them at scale} %
{will in many cases require} %
{technical resources beyond the means of all but the biggest providers. As a result, whether through their own AI services or through those of others who depend on their cloud infrastructure, major providers like AWS, Microsoft Azure, and Google Cloud will be crucial players in future AI development and distribution}. 

Some AI-specific providers' services can be accessed only through a larger provider's interface and brought by customers into applications through that specific provider’s cloud, rather than through a competitor (OpenAI’s commercial services can be accessed \textit{only} through Azure~\cite{msopenai}). The larger provider’s cloud offering thus operates as a platform through which they facilitate and can gatekeep market access to the smaller provider’s service. %
{In some cases, one cloud provider's interface may be used to access a specialist AI provider's model~\cite{awsintermed}, where that specialist provider itself uses a \textit{different} cloud provider for their supporting infrastructure for development~\cite{anthropicgcloud}. That is to say, several larger cloud providers may be involved at different stages of production and deployment of specialist AI providers' services {(and indeed, those of others).}} %

\subsection{Supply chains are increasingly consolidating around systemically important providers}
\label{sec:consolidation}

The dynamics of interdependence and integration mean that algorithmic supply chains are increasingly consolidating around (primarily) Amazon, Microsoft, and Google~\cite{cobbe2021artificial,cobbe2019whatlies,statista}. Several factors tend towards consolidation, including competitive advantages offered by integration. These companies span markets, offering developers `all-in-one’ packages %
with easy access to state-of-the-art technologies, which readily scale and enable `global' reach. In AI production, they leverage bespoke and advanced computing resources and expertise, significant quantities of data representing real-world deployments and use-cases, and economies of scale across AI and non-AI customer bases. They can therefore offer %
services at lower cost, broader scale, greater technical sophistication, and with potentially easier access %
than many competitors. Moreover, their substantial financial resources help consolidate their position through purchases of and investments in potential competitors (such as Google’s purchase of DeepMind~\cite{guardiandeepmind}, or Microsoft’s investment in OpenAI~\cite{msopenai}).

As a result, %
major %
providers are \textit{systemically important} for the political economy, governance, and accountability of AI. Even where an application does not use a major provider's AI services (%
using the developer's own AI technology, for example, or obtain%
ing it from a smaller provider), major providers' non-AI services may form significant parts of %
supply chains for either that application or the AI service it uses (or both). These providers are therefore core actors in many supply chains%
, strategically positioning themselves across markets %
in a process of enclosure of AI-technological infrastructure and, by extension, of businesses, institutions, organisations, and sectors %
relying on supply chains involving their services. They are thus positioned in commercially beneficial interdependencies both %
with other actors in particular supply chains, but also %
\changejc{more broadly} -- a few dominant providers underpin important social and economic processes while themselves depending {to various degrees} on many %
actors in %
social, legal, technological, and political economic processes which help produce and maintain their position. %

%% file: s4-implications.tex
\section{(Implications for) Accountability in algorithmic supply chains}
\label{sec:supplychainaccountability}

Algorithmic supply chains bring difficult implications for governance and accountability. Much algorithmic accountability research often reflects an organisation-focused %
understanding of accountability (\S\ref{sec:algosystemaccountability}). Yet the production, deployment, and use of AI technologies in supply chains is split between multiple %
actors who together produce its workings and effects and whose part in producing functionality cannot be understood separately from the chain (\S\ref{sec:interdependence}). Organisation-focused framings cannot properly capture this distribution of responsibilities between actors across the stages of the AI lifecycle, which also challenges assignments of accountability in relevant legal frameworks (\S\ref{sec:distributedresponsibility}). Moreover, problems with systems can propagate widely downstream through supply chains (\S\ref{sec:propagation}), yet particular actors are often unaware of the broader chain, and the %
limits of visibility across supply chains make interventions like risk assessments difficult (\S\ref{sec:accountabilityhorizon}). %
It is therefore crucial for governance and accountability mechanisms to understand the actors in supply chains, what they do for each other, which of them take core roles, and the interconnections and interdependencies between them. At the same time, however, the dynamic, transient nature of supply chains (\S\ref{sec:process})---which can potentially be instantiated each time and unfold differently as data is processed---is also challenging.

Moreover, algorithmic supply chains are structured through %
interactions between technological, legal, social, and political economic factors  (\S\ref{sec:structuringinterdependence}). It is therefore not enough to attend only to ways of making the technology more transparent or understandable 
(though this can help understand specific points in particular systems' lifecycle). Instead, algorithmic accountability work must %
{consider} %
broader factors: %
how providers leverage technology and law to structure interdependencies, integrate their operations (\S\ref{sec:integration}), consolidate their position (\S\ref{sec:consolidation}), increase their control and power while minimising legal accountability (\S\ref{sec:servitisation}), and extend their supply chains across borders to minimise cost and legal risk and maximise commercial benefit ({\S\ref{sec:arbitrage}}). The dynamics of supply chains, the legal and political economic factors influencing their structure, and the relations and interdependencies between actors that result are all significant considerations from a view of accountability \textit{as a mechanism}---one, in particular, for investigating, understanding, assessing, challenging, and contesting power. They are also important in considering who should be accountable, to whom, for what, and through which mechanisms and institutional arrangements %

\subsection{Responsibility for algorithmic systems is distributed between several actors}
\label{sec:distributedresponsibility}

Governance and accountability mechanisms around algorithmic systems should address the \textit{distributed responsibility} in algorithmic supply chains. Different actors control aspects of commissioning, designing, developing, deploying, using, or monitoring a particular AI technology.  %
Responsibility for the workings and outcomes of supply chains is thus distributed among several actors who may not be straightforward to identify nor consistent across instances. Even when some actors are influential, there is therefore typically no one actor in overall control of a supply chain. Existing accountability literature, however, typically assumes that (while models or input data might change) the actors and components %
remain relatively stable. %
Yet directing governance and accountability mechanisms at, or allocating accountability to, the wrong actors in supply chains risks undermining the stated goals of these mechanisms. %
\changejc{Accountability involves a relationship where an actor provides accounts of their activities to a forum, who imposes consequences to correct the actor if needed~\cite{bovens2006}. For accountability mechanisms to succeed, it is therefore crucial that the right actors are assigned to the appropriate relationships. In this context, those who are factually responsible for various aspects of production, distribution, and use of algorithmic systems must be identified correctly so that 
accountability can be allocated accordingly.
}
\subsubsection{Legal accountability and distributed responsibility}

Some %
\changejc{jurisdictions} have sought to address distributed responsibility in data-driven supply chains more generally. The Court of Justice of the European Union (CJEU) has attempted to contend with this in data protection law, for example. A key question in data protection law is who is a \textit{data controller} -- factually in control of, and therefore primarily responsible in law for, personal data processing~\cite{EU-GDPR}. The CJEU has repeatedly held that multiple parties can be controllers for some or all aspects of a chain of processing~\cite{mahieu2019,cobbe2021artificial,jehovan,fashionid,wirtschaftsakademie}. Where several actors have common interests in the %
processing, they may be \textit{joint controllers}~\cite{EU-GDPR}; where their interests in the processing diverge, they may be \textit{separate controllers}. In doing so, the Court made several %
observations: actors can be controllers if they have influence despite not having actual access to the personal data~\cite{wirtschaftsakademie,jehovan,fashionid}, controllers are not typically responsible for parts of the chain %
\changejc{before or after} those they actually influence \cite{fashionid}; %
and using another actor’s platform does not exempt a controller from their obligations~\cite{wirtschaftsakademie}.

Recognising the plurality of actors in chains of processing is welcome, but even data protection law's more nuanced assignment of roles and responsibilities may not readily map to algorithmic supply chains~\cite{gurses2017,cobbe2021artificial,mahieu2019}. Under current understandings, AI service customers are likely data controllers (the dominant party, ultimately responsible for compliance and accountability), while providers may be \textit{data processors} (the subordinate party, acting only under the instruction of a controller, with limited obligations)~\cite{millard2021,cobbe2021artificial,mahieu2019}. Yet this assignment of legal roles and responsibilities does not describe the real interdependencies and power relations between AI service providers (who are in control of their technologies, often core actors in supply chains, potentially systemically important more generally, typically presenting customers with `take-it-or-leave-it' contracts, and to a large extent determining AI-driven functionality in customers’ applications %
through their production processes) and their customers (potentially small companies without AI expertise, typically with no access to the provider’s systems, control over them, or knowledge of how they work)~\cite{cobbe2021artificial}. Even  where providers \textit{are} likely controllers for aspects of the service---such as where they use customer data for service improvement---they typically attempt to minimise responsibility by claiming in their service agreements to be processors \cite{cobbe2021artificial}. Yet the CJEU has consistently held that the factual situation outweighs contractual or other arrangements, and regulators have contradicted claimed assignments of legal roles in other kinds of data-driven supply chains~\cite{vealeImpossibleAsksCan2022}. Challenging providers' claims, however, would involve litigation or regulatory investigation. Moreover, given the need for joint controllers to agree the division of controllers' duties and responsibilities between themselves~\cite{EU-GDPR}, it is not clear how joint controllership can work where actors don't necessarily know of each other or have any direct relationship. 

The EU’s proposed AI Act suffers from related tensions. It recognises that the ‘user’ %
of an AI system {(in this context, generally the customer of an AI service)} may differ from its ‘provider’, and envisages circumstances where a user of an AI service does so for a purpose not intended by the provider, and thus in law becomes responsible for %
{the underlying system}~\cite{EU-AIAct}. However, this does not reflect supply chain interdependencies and dynamics, where production, deployment, and use are distributed between actors. Instead, in this circumstance, the Act would potentially make actors several steps downstream from production responsible for ensuring that the AI technology complies with production-related legal requirements around training and testing, accountability, and risk management. While the user %
would inevitably be unable to comply {(due to the actual distribution of practical responsibilities in algorithmic supply chains)}, the actor who developed and controls that technology and is thus factually responsible for production would face no obligations. This may incentivise actors who can never provide assurance of compliance to pretend they can -- easily done due to the Act's self-regulatory framework and limited planned regulatory capacity~\cite{vealeDemystifyingDraftEU2021}. Regulatory systems which hold supply chain actors downstream of production to account for design and development may do little to regulate those who are factually responsible for production and who benefit financially from potentially unlawful API queries, effectively shielding them from liability.

\subsubsection{Allocating accountability}

Governance and accountability mechanisms should therefore be grounded more clearly in and emphasise an understanding of the distribution of responsibility in algorithmic supply chains. Not \textit{every} actor in a supply chain will be responsible for the outcome of the algorithmic system -- some will provide only supporting services which do not meaningfully affect outcomes. Neither will actors who \textit{are} in some way responsible be \textit{equally} responsible -- some play a bigger role than others in determining outcomes. Nor will they be responsible for the \textit{whole} supply chain -- different actors control different aspects of it. Accountability should thus be allocated to actors across supply chains based on a proper understanding of their technological and political economic dynamics. This requires processes and criteria for identifying the distribution of responsibility across supply chains and allocating accountability to those actors, for which activities, accounting to whom, and with what possible consequences. 

It %
is therefore important to understand the distribution of responsibility in algorithmic supply chains in terms of who is doing what for whom, who is performing what key functions for others, who is core to certain supply chains, and who is systemically important. Particular attention is %
due to systemically important actors -- %
primarily Amazon, Microsoft, Google, and perhaps a few others. Though technological and political economic dynamics tend towards consolidation around these companies, and though non-AI services often provide supporting infrastructure, it %
\changejc{is} still %
important to have ways to determine which aspects of supply chains are key to their outcomes and effects, as opposed to those which  could be interchanged without affecting those things. The latter, while potentially significant%
, are perhaps less of an urgent subject of governance and accountability mechanisms than the former.

\subsection{The \textit{accountability horizon} limits visibility across supply chains}
\label{sec:accountabilityhorizon}

A significant challenge for governance and accountability mechanisms in algorithmic supply chains is the \textbf{accountability horizon} -- the point beyond which an actor cannot ‘see’, which depends on the actor and the %
chain. Supply chain actors will generally be able to 
know whom they interact with directly (a first ‘step’ in the chain), and perhaps whom those first step actors interact with in turn (a second ‘step’), but may not %
\changejc{be able to know about} the data flows and interconnections beyond~\cite{cobbe2019whatlies,singh2019decprov}. Moreover, distributed responsibility %
between actors (\S\ref{sec:distributedresponsibility}) means  each has incomplete information even if they \textit{do} know who is up- and downstream of them. AI service providers, in control of production, may %
lack knowledge of downstream contexts and use-cases of application deployments~\cite{lewicki2023aiaasfairness}. Those responsible for deployment and use typically lack access to models and often to information about their specification, training, testing, validation, and so on (and thus may have limited understanding of their capabilities and limitations). 

The accountability horizon is thus a problem for producers of algorithmic systems in the earliest stages of developing their technologies (\textit{problem framing}, 
 \S\ref{sec:problemframing}) and for legal and other governance frameworks based around \textit{risk management} (\S\ref{sec:riskmanagement}).

\subsubsection{The accountability horizon makes problem framing difficult}
\label{sec:problemframing}

Many algorithmic issues stem from choices around problem definition and framing that inform system design. Complex concepts may be formalised poorly, tasks may be incompletely captured, and different contexts may be insufficiently considered \cite{selbstFairnessAbstractionSociotechnical2019}. The `many hands' problem made critical questions of identifying who framed the problem and when it was framed difficult to answer \cite{passiProblemFormulationFairness2019}. Supply chain dynamics \textit{giving rise to the accountability horizon} complicate this further. Those responsible for production have limited capacity to understand the contexts of deployment and use by others, while the actors closest to the problem---those deploying or using the system---are generally unable to influence its design. Moreover, due to the split between production and deployment, application developers necessarily engage in their own problem framing -- determining whether they need an AI service to address a particular problem and, if so, which is most suitable. Yet they may lack capacity to determine which service (if any) is most appropriate to their needs (particularly if organisations swap organisational and IT know-how for license managers~\cite{balaynDebiasingRegulatingAI2021}). This is further complicated by the fact that not all services are fungible, or adaptable to a range of different framings. Services may only accept certain kinds of input data, produce certain kinds of output data, or be amenable to certain kinds of alteration and customisation. They may be developed with particular underlying assumptions which can (or should) preclude their deployment or use in other contexts~\cite{lewicki2023aiaasfairness}. Supply chain integration may further reduce flexibility in problem framing, as technical hurdles to limit interoperability and cost implications make components less readily swappable (particularly where services are strategically bundled by providers).

More cynically, actors may encourage problem framing which increases demand for their own products and services. For example, organisations selling technologies for input data, such as cameras and other environmental sensors, may also sell workplace monitoring tools which take advantage of the data produced by these sensors. Application developers with low problem framing capacity might adopt these tools without properly identifying whether they need workplace monitoring at all. The dependency of application developers on supply chains may therefore risk the autonomy of those organisations~\cite{balaynDebiasingRegulatingAI2021}. Indeed, using 
AI services leaves healthcare, education and other established sectors vulnerable to unbundling and rebundling of their  fundamental operations, leaving each stage amenable to value extraction through servitisation~\cite{balaynDebiasingRegulatingAI2021}. %

\subsubsection{The accountability horizon makes risk management difficult}
\label{sec:riskmanagement}

Many academic, policy, and legislative initiatives %
\changejc{propose} impact assessments, risk assessments, and risk management mechanisms to mitigate harms %
\changejc{of} AI technologies (for example, ~\cite{EU-AIAct,cobbe2021reviewable,javadi2020monitoring,edwards2017slave,koene2019,krollaa,hutchinson2021towards,raji2020closing,williams2021,kaminski2021,rader2018explanations,ico2020,arrieta2020,canadaAlgAssess,ainow,lee2021risk,krafft2021action,saleiro2018aequitas,mitchell2019model,gebru2018datasheets,metaxa2021auditing,demos2020,adalovelace2020}). {Data controllers' management of risks to data subjects' fundamental rights is also core %
to the EU's data protection regime~\cite{EU-GDPR,cobbeDPDW}. %
Typically, certain actors---who may differ between legal frameworks---have some obligation to identify and mitigate risks to individuals or their rights %
arising from technologies they develop or control.}

However, the accountability horizon makes %
effective risk management difficult if not impossible%
. These %
mechanisms require knowledge of both the AI technology's specification and development (i.e. production) \textit{and} the purpose and context of its application (deployment)~\cite{cobbe2021artificial,lewicki2023aiaasfairness}.  %
{Yet without %
\changejc{advance knowledge of} their customers' \changejc{many, varied, and changing} application contexts and uses%
, providers cannot properly account for the range of potential risks that might arise~{\cite{cobbe2021artificial,javadi2020monitoring}. Similarly, without knowledge of or influence over production, customers cannot reliably assess how systems are developed, nor %
ensure that systems are appropriate to the risks arising in their context~\cite{lewicki2023aiaasfairness}.} %
Even where they have some knowledge, models are regularly updated, and %
{customers} may lack visibility or capacity to reassess. In many cases, therefore, no actor will have sufficient knowledge of or control over both production and deployment to be able to reliably assess or mitigate the impacts and risks. %
Risk management approaches to governance and accountability of AI technologies are {therefore arguably not} appropriate in this context (despite their importance in emerging laws applying to algorithmic systems, such as the EU’s AI Act~\cite{EU-AIAct}).%

\subsubsection{Expanding the accountability horizon}

The accountability horizon thus poses major problems for accountability. 
Interventions %
{are needed to} %
expand the %
horizon and better place actors to 
know more about their own supply chains, and %
support others in knowing more about theirs. 
Yet organisation-focused tools to provide information on points in the AI lifecycle (such as ~\cite{gebru2018datasheets,mitchell2019model}) are of limited help where information about interconnections between actors is needed. %
Tracking data flow between actors could help understand interconnections beyond the first few steps~\cite{singh2019decprov}, as could legal and institutional mechanisms requiring information about arrangements. 
`Know your customer’ requirements around %
on-boarding for AI services (common in financial services) could help providers understand customers’ purposes and intentions~\cite{cobbe2021artificial}\changejs{, as can technical measures for monitoring how their services are (mis)used~\cite{javadi2020monitoring,javadi2021monitoring}} 
(though these only give 
some visibility over one or two steps in the chain). %
Recent CJEU data protection jurisprudence on transparency rights confirms that data subjects have the right to know the %
\changejc{identities of} recipients of their personal data~\cite{osterreichische}, which may help %
understand data flows. However, where the data controller does not know %
\changejc{those} identities%
\changejs{---perhaps likely}
due to the accountability horizon---data subjects can instead be told about the \textit{categories} of recipients%
~\cite{osterreichische} (significantly less useful information). %

A particular difficulty{, however,} is that accountability is contextual~\cite{cobbe2021reviewable,williams2021,wieringa2020account,bovens2006,AnannyCrawford_2018,norval2022disclosures}. The information needed to account for an algorithmic system depends on the actors responsible for its development, deployment, and use, on the forum owed the account, and on the broader context~\cite{cobbe2021reviewable}. As such, the mechanisms needed to record, process, and provide information about algorithmic systems---such as ~\cite{gebru2018datasheets,mitchell2019model,arnold2019factsheets,kroll2021traceability,cloete2021droiditor,crisan2022interactivemc}---are 
also contextual. %
Yet the accountability horizon makes understanding context difficult for those who account for their part in %
supply chains. They may not know whom they need to account to, so may not know what information to retain, about what aspects of their processes, and in what form.
They may also not know which actors upstream from them they can obtain accounts from.
In general, the difficulties %
raised by the accountability horizon are not easily overcome. 

\subsection{Providers structure supply chains  to minimise accountability}%
\label{sec:benefit}

Supply chain dynamics allow providers to maximise commercial benefit while minimising legal accountability by \textit{(i)} extending control over downstream deployment of AI technologies (\S\ref{sec:servitisation}) and \textit{(ii)} extending their own supply chains across borders to engage in regulatory arbitrage (\S\ref{sec:arbitrage}). Providers thus use a %
techno-legal strategy to position themselves advantageously in markets and shape supply chains to maximise revenue and reduce risk~\cite{cohenPiracySecurityArchitectures2012,cohen2019}.

\subsubsection{Servitised distribution models give providers control beyond deployment%
}
\label{sec:servitisation}

Nissenbaum observed that%
, in the mid-1990s, %
software vendors %
often demanded property protection for their products while denying, as far as possible, accountability for them~\cite{nissenbaum1996}. Software %
licensing agreements precluded \textit{ownership} by %
users %
and emphasised the producer’s rights, while disclaiming their legal accountability for the software or anything it might do -- even where harms resulted directly from defects in it~\cite{nissenbaum1996}. %
Developers thus attempted to maintain control over their software to the extent possible given the distribution model at the time (typically physical media), while generating artificial scarcity for an information product to maximise revenue with minimal risk and responsibility. This produced, as Nissenbaum puts it, a ‘vacuum’ of accountability~\cite{nissenbaum1996}.

Software's distribution \cut{model}has \changejc{moved} away from (licensed) physical media to the
service-based, API-centric models described \cut{above}~\cite{gurses2017}. %
Combined with asymmetrical interdependence in algorithmic supply chains (\S\ref{sec:coreandperiphery}), %
service %
models offer providers new %
ways to extend control past the point of deployment. \cut{In particular, b}\changejs{B}ecause providers depend less on individual customers than %
customers %
depend on them, providers can %
impose 
contractual service agreements and use APIs as tools %
to advantageously structure their relations with customers and others. %
Where vendors once sought expansive intellectual property protections,\cut{for instance, %
providers today}
\changejs{today providers} seek to {use %
service agreements to} maximise control over {deployment of} their technologies by reserving rights to dictate terms of use and change, withdraw, or cancel products and services at will. %
Providers disclaim legal accountability for things that happen through use of their services~\cite{kamarinou2015,turton2021,michelsStandardContractsCloud2021}, and attempt to position themselves as data processors (\S\ref{sec:distributedresponsibility}) even \changejs{when}\cut{where} %
using customer data for their own purposes and %
thus \changejs{are} likely %
the controller for that processing~\cite{cobbe2021artificial}. %
{Providers can also use} APIs %
as `projections'~\cite{bucherObjectsIntenseFeeling2013} of the asymmetric balances of power %
{with} customers, to destabilise attempts to hold %
{providers} to account: using APIs as tools to shutter businesses, undermine research, %
and evade scrutiny~\cite{ausloosResearchingDataRights2020}, while contractually giving themselves those rights and %
\changejc{using} changes in information policy \cut{as a form of}\changejs{to} control~\cite{reidenbergLexInformaticaFormulation1998}.

\subsubsection{Cross-border supply chains permit regulatory arbitrage}
\label{sec:arbitrage}

As we describe, data processing and networking technologies afford a geographical distribution of AI production and deployment  (\S\ref{sec:structuringinterdependence}). The same technologies allow various production-related activities to themselves be distributed geographically, incentivised by jurisdictional differences in cost and regulation. This allows regulatory arbitrage, where companies in one jurisdiction exploit legal and political economic conditions in other jurisdictions to maximise commercial benefit while minimising legal accountability. This often involves contracting third-parties (such as Sama AI~\cite{samaAI} {or Supahands~\cite{supahands}}%
) to undertake some aspects of production%
. For example, differences in privacy and data protection laws and labour protections can lower the %
legal risk of dataset production activities like data %
cleaning, and labelling~\cite{matusCertificationSystemsMachine2022,grayGhostWorkHow2019,timeOpenAI}. Environmental factors like cheap water and energy and lax planning and waste laws can influence the location of compute and storage~\cite{moscoDarkClouds2014}. 

While some laws---such as the EU's data protection law~\cite{EU-GDPR} and AI Act~\cite{EU-AIAct} and California's Consumer Privacy Act~\cite{US-CCPA}---have sought extra-territorial effect to address regulatory arbitrage, the cross-border nature of supply chains and difficulties of enforcement remains a significant accountability challenge.

%% file: s5-conclusion.tex
\section{Conclusions and further research}
\label{sec:conclusion}

The `many hands' problem has motivated efforts to provide information about the production, deployment, and use of algorithmic systems by teams and organisations (\S\ref{sec:algosystemaccountability}). The emergence of AI `as a service' {(or `general purpose AI')} and developments associated with cloud computing and the services model of software distribution (\S\ref{sec:supplychains}) challenge organisation-focused understandings of algorithmic accountability (\S\ref{sec:supplychainaccountability}) in ways that have not been widely addressed%
. 

AI technologies now often involve \textit{algorithmic supply chains}, with their production, deployment, and use split between multiple actors who \textit{together} produce the technology's outcomes and functionality (\S\ref{sec:interdependence}). Major providers---now highly integrated {both horizontally and vertically} %
(\S\ref{sec:integration})---are systemically important players (\S\ref{sec:coreandperiphery}), and supply chains are increasingly consolidating around them (\S\ref{sec:consolidation}). Issues with particular systems can propagate through supply chains (\S\ref{sec:propagation}), while they often change between instances, making it difficult to understand how they operate or who is involved (\S\ref{sec:process}). Together, these dynamics of interdependence, perpetual change, integration, and consolidation produce supply chains in which responsibility for algorithmic systems is distributed between interdependent actors (\S\ref{sec:distributedresponsibility}) and visibility across the actors involved is low (\S\ref{sec:accountabilityhorizon}). This challenges existing legal accountability frameworks while limiting the effectiveness of mechanisms like risk assessments. Moreover, splitting production and deployment makes it difficult to appropriately develop or choose AI services (\S\ref{sec:problemframing}). At the same time, the services distribution model allows providers to use terms of service and APIs to minimise legal accountability and maximise control over technologies beyond deployment (\S\ref{sec:servitisation}), while simultaneously extending their own production processes across borders to exploit differences in regulatory regimes (\S\ref{sec:arbitrage}).

In all, the characteristics of algorithmic supply chains we have identified and the implications they raise challenge existing approaches to algorithmic accountability. Future algorithmic accountability research must therefore contend with supply chain dynamics: how they are structured, how they develop over time, how AI's functionality and effects are produced through them{, and---importantly---how distributed responsibility challenges governance mechanisms and the accountability horizon limits visibility.} %
This requires a broad view of %
supply chains, seeking to understand %
who is involved, what they are doing, and how to allocate accountability between them. Importantly, supply chains are structured by legal and political economic factors, which must be properly understood, as well as technological ones. If governance and accountability mechanisms are to hold those responsible for developing, deploying, and using AI technologies to account for their workings and effects, the dynamics of supply chains must be urgently addressed.

%% file: main.bbl
\begin{thebibliography}{110}


\ifx \showCODEN    \undefined \def \showCODEN     #1{\unskip}     \fi
\ifx \showDOI      \undefined \def \showDOI       #1{#1}\fi
\ifx \showISBNx    \undefined \def \showISBNx     #1{\unskip}     \fi
\ifx \showISBNxiii \undefined \def \showISBNxiii  #1{\unskip}     \fi
\ifx \showISSN     \undefined \def \showISSN      #1{\unskip}     \fi
\ifx \showLCCN     \undefined \def \showLCCN      #1{\unskip}     \fi
\ifx \shownote     \undefined \def \shownote      #1{#1}          \fi
\ifx \showarticletitle \undefined \def \showarticletitle #1{#1}   \fi
\ifx \showURL      \undefined \def \showURL       {\relax}        \fi
\providecommand\bibfield[2]{#2}
\providecommand\bibinfo[2]{#2}
\providecommand\natexlab[1]{#1}
\providecommand\showeprint[2][]{arXiv:#2}

\bibitem[{Ada Lovelace Institute}(2020)]%
        {adalovelace2020}
\bibfield{author}{\bibinfo{person}{{Ada Lovelace Institute}}.}
  \bibinfo{year}{2020}\natexlab{}.
\newblock \bibinfo{title}{{Examining the Black Box: Tools for Assessing
  Algorithmic Systems}}.
\newblock
\newblock
\urldef\tempurl%
\url{https://www.adalovelaceinstitute.org/report/examining-the-black-box-tools-for-assessing-algorithmic-systems/}
\showURL{%
\tempurl}


\bibitem[{Amazon}(n.\,d.)]%
        {mturk}
\bibfield{author}{\bibinfo{person}{{Amazon}}.}
  \bibinfo{year}{[n.\,d.]}\natexlab{}.
\newblock \bibinfo{title}{{Mechanical Turk}}.
\newblock \bibinfo{howpublished}{https://www.mturk.com}.
\newblock



\bibitem[Amazon(2021)]%
        {AmazonAIServices}
\bibfield{author}{\bibinfo{person}{Amazon}.} \bibinfo{year}{2021}\natexlab{}.
\newblock \bibinfo{title}{Artificial Intelligence Services}.
\newblock
  \bibinfo{howpublished}{https://aws.amazon.com/ machine-learning/ai-services}.
\newblock
\newblock
\shownote{(Accessed on 04/11/2021)}.


\bibitem[{Amazon}(2021)]%
        {amazonCaseStudyWeShine2021}
\bibfield{author}{\bibinfo{person}{{Amazon}}.} \bibinfo{year}{2021}\natexlab{}.
\newblock \bibinfo{title}{Case {{Study}}: {{WeShine}}}.
\newblock \bibinfo{howpublished}{https://perma.cc/2V2T-NNKT}.
\newblock


\bibitem[Ananny and Crawford(2018)]%
        {AnannyCrawford_2018}
\bibfield{author}{\bibinfo{person}{Mike Ananny} {and} \bibinfo{person}{Kate
  Crawford}.} \bibinfo{year}{2018}\natexlab{}.
\newblock \showarticletitle{Seeing without knowing: Limitations of the
  transparency ideal and its application to algorithmic accountability}.
\newblock \bibinfo{journal}{\emph{New Media and Society}} \bibinfo{volume}{20},
  \bibinfo{number}{3} (\bibinfo{year}{2018}), \bibinfo{pages}{973–989}.
\newblock
\showISSN{14617315}
\urldef\tempurl%
\url{https://doi.org/10.1177/1461444816676645}
\showDOI{\tempurl}


\bibitem[Anthropic(2023)]%
        {anthropicgcloud}
\bibfield{author}{\bibinfo{person}{Anthropic}.}
  \bibinfo{year}{2023}\natexlab{}.
\newblock \bibinfo{title}{{Anthropic Partners with Google Cloud}}.
\newblock
  \bibinfo{howpublished}{https://www.anthropic.com/index/anthropic-partners-with-google-cloud-2}.
\newblock


\bibitem[Arnold et~al\mbox{.}(2019)]%
        {arnold2019factsheets}
\bibfield{author}{\bibinfo{person}{Matthew Arnold}, \bibinfo{person}{Rachel~KE
  Bellamy}, \bibinfo{person}{Michael Hind}, \bibinfo{person}{Stephanie Houde},
  \bibinfo{person}{Sameep Mehta}, \bibinfo{person}{Aleksandra Mojsilovi{\'c}},
  \bibinfo{person}{Ravi Nair}, \bibinfo{person}{K~Natesan Ramamurthy},
  \bibinfo{person}{Alexandra Olteanu}, \bibinfo{person}{David Piorkowski},
  {et~al\mbox{.}}} \bibinfo{year}{2019}\natexlab{}.
\newblock \showarticletitle{FactSheets: Increasing trust in AI services through
  supplier's declarations of conformity}.
\newblock \bibinfo{journal}{\emph{IBM Journal of Research and Development}}
  \bibinfo{volume}{63}, \bibinfo{number}{4/5} (\bibinfo{year}{2019}),
  \bibinfo{pages}{6--1}.
\newblock


\bibitem[Arrieta et~al\mbox{.}(2020)]%
        {arrieta2020}
\bibfield{author}{\bibinfo{person}{Alejandro~Barredo Arrieta} {et~al\mbox{.}}}
  \bibinfo{year}{2020}\natexlab{}.
\newblock \showarticletitle{Explainable Artificial Intelligence (XAI):
  Concepts, Taxonomies, Opportunities and Challenges toward Responsible AI}.
\newblock \bibinfo{journal}{\emph{Information Fusion}}  \bibinfo{volume}{58}
  (\bibinfo{year}{2020}).
\newblock


\bibitem[Ausloos and Veale(2020)]%
        {ausloosResearchingDataRights2020}
\bibfield{author}{\bibinfo{person}{Jef Ausloos} {and} \bibinfo{person}{Michael
  Veale}.} \bibinfo{year}{2020}\natexlab{}.
\newblock \showarticletitle{Researching with {{Data Rights}}}.
\newblock \bibinfo{journal}{\emph{Technology and Regulation}}
  \bibinfo{volume}{2020}, \bibinfo{number}{1} (\bibinfo{year}{2020}),
  \bibinfo{pages}{136--157}.
\newblock


\bibitem[Balayn and G{\"u}rses(2021)]%
        {balaynDebiasingRegulatingAI2021}
\bibfield{author}{\bibinfo{person}{Agathe Balayn} {and} \bibinfo{person}{Seda
  G{\"u}rses}.} \bibinfo{year}{2021}\natexlab{}.
\newblock \bibinfo{title}{Beyond Debiasing: {{Regulating AI}} and Its
  Inequalities}.
\newblock \bibinfo{howpublished}{https://perma.cc/4UAV-3UFB}.
\newblock


\bibitem[Bovens(1998)]%
        {bovens1998}
\bibfield{author}{\bibinfo{person}{Mark Bovens}.}
  \bibinfo{year}{1998}\natexlab{}.
\newblock \bibinfo{booktitle}{\emph{The quest for responsibility.
  Accountability and citizenship in complex organisations}}.
\newblock \bibinfo{publisher}{Cambridge University Press}.
\newblock


\bibitem[Bovens(2006)]%
        {bovens2006}
\bibfield{author}{\bibinfo{person}{Mark Bovens}.}
  \bibinfo{year}{2006}\natexlab{}.
\newblock \showarticletitle{Analysing and Assessing Public Accountability. A
  Conceptual Framework}.
\newblock \bibinfo{journal}{\emph{EUROGOV, European Governance Papers}}
  \bibinfo{volume}{No. C-06-01} (\bibinfo{year}{2006}).
\newblock


\bibitem[Bovens(2010)]%
        {bovens2010}
\bibfield{author}{\bibinfo{person}{Mark Bovens}.}
  \bibinfo{year}{2010}\natexlab{}.
\newblock \showarticletitle{Two Concepts of Accountability: Accountability as a
  Virtue and as a Mechanism}.
\newblock \bibinfo{journal}{\emph{West European Politics}}
  \bibinfo{volume}{33}, \bibinfo{number}{5} (\bibinfo{year}{2010}),
  \bibinfo{pages}{946--967}.
\newblock
\urldef\tempurl%
\url{https://doi.org/10.1080/01402382.2010.486119}
\showDOI{\tempurl}


\bibitem[Bucher(2013)]%
        {bucherObjectsIntenseFeeling2013}
\bibfield{author}{\bibinfo{person}{Tania Bucher}.}
  \bibinfo{year}{2013}\natexlab{}.
\newblock \showarticletitle{Objects of {{Intense Feeling}}: {{The Case}} of the
  {{Twitter API}}}.
\newblock \bibinfo{journal}{\emph{Computational Culture: A Journal of Software
  Studies}}  \bibinfo{volume}{3} (\bibinfo{year}{2013}).
\newblock


\bibitem[Buolamwini and Gebru(2018)]%
        {buolamwini2018gender}
\bibfield{author}{\bibinfo{person}{Joy Buolamwini} {and}
  \bibinfo{person}{Timnit Gebru}.} \bibinfo{year}{2018}\natexlab{}.
\newblock \showarticletitle{Gender shades: Intersectional accuracy disparities
  in commercial gender classification}. In \bibinfo{booktitle}{\emph{Conference
  on Fairness, Accountability, and Transparency}}. PMLR,
  \bibinfo{pages}{77--91}.
\newblock


\bibitem[{CJEU}(2018a)]%
        {wirtschaftsakademie}
\bibfield{author}{\bibinfo{person}{{CJEU}}.} \bibinfo{year}{2018}\natexlab{a}.
\newblock \bibinfo{title}{Case C‑210/16, Unabh\"{a}ngiges Landeszentrum
  f\"{u}r Datenschutz Schleswig-Holstein v Wirtschaftsakademie
  Schleswig-Holstein GmbH}.
\newblock \bibinfo{howpublished}{ECLI:EU:C:2018:388}.
\newblock


\bibitem[{CJEU}(2018b)]%
        {jehovan}
\bibfield{author}{\bibinfo{person}{{CJEU}}.} \bibinfo{year}{2018}\natexlab{b}.
\newblock \bibinfo{title}{Case C‑25/17, Jehovan todistajat}.
\newblock \bibinfo{howpublished}{ECLI:EU:C:2018:551}.
\newblock


\bibitem[{CJEU}(2019)]%
        {fashionid}
\bibfield{author}{\bibinfo{person}{{CJEU}}.} \bibinfo{year}{2019}\natexlab{}.
\newblock \bibinfo{title}{Case C-49/17, Fashion ID GmbH \& Co.KG v
  Verbraucherzentrale NRW eV.}
\newblock \bibinfo{howpublished}{ECLI:EU:C:2019:629}.
\newblock


\bibitem[{CJEU}(2023)]%
        {osterreichische}
\bibfield{author}{\bibinfo{person}{{CJEU}}.} \bibinfo{year}{2023}\natexlab{}.
\newblock \bibinfo{title}{{Case C-154/21, Österreichische Post}}.
\newblock \bibinfo{howpublished}{ECLI:EU:C:2023:3}.
\newblock


\bibitem[{Clearview AI}(n.\,d.)]%
        {clearviewAI}
\bibfield{author}{\bibinfo{person}{{Clearview AI}}.}
  \bibinfo{year}{[n.\,d.]}\natexlab{}.
\newblock \bibinfo{title}{{https://www.clearview.ai}}.
\newblock \bibinfo{howpublished}{\url{https://www.clearview.ai}}.
\newblock


\bibitem[Cloete et~al\mbox{.}(2022)]%
        {cloete2021droiditor}
\bibfield{author}{\bibinfo{person}{Richard Cloete}, \bibinfo{person}{Chris
  Norval}, {and} \bibinfo{person}{Jatinder Singh}.}
  \bibinfo{year}{2022}\natexlab{}.
\newblock \showarticletitle{Auditable Augmented/Mixed/Virtual Reality: The
  Practicalities of Mobile System Transparency}.
\newblock \bibinfo{journal}{\emph{Proc. ACM Interact. Mob. Wearable Ubiquitous
  Technol.}} \bibinfo{volume}{5}, \bibinfo{number}{4}, Article
  \bibinfo{articleno}{149} (\bibinfo{date}{dec} \bibinfo{year}{2022}),
  \bibinfo{numpages}{24}~pages.
\newblock
\urldef\tempurl%
\url{https://doi.org/10.1145/3495001}
\showDOI{\tempurl}


\bibitem[Cobbe et~al\mbox{.}(2019)]%
        {cobbe2019whatlies}
\bibfield{author}{\bibinfo{person}{Jennifer Cobbe} {et~al\mbox{.}}}
  \bibinfo{year}{2019}\natexlab{}.
\newblock \showarticletitle{What lies beneath: Transparency in online service
  supply chains}.
\newblock \bibinfo{journal}{\emph{Journal of Cyber Policy}}
  \bibinfo{volume}{5}, \bibinfo{number}{1} (\bibinfo{year}{2019}).
\newblock


\bibitem[Cobbe et~al\mbox{.}(2021)]%
        {cobbe2021reviewable}
\bibfield{author}{\bibinfo{person}{Jennifer Cobbe}, \bibinfo{person}{Michelle
  Seng~Ah Lee}, {and} \bibinfo{person}{Jatinder Singh}.}
  \bibinfo{year}{2021}\natexlab{}.
\newblock \showarticletitle{Reviewable Automated Decision-Making: A Framework
  for Accountable Algorithmic Systems}. In
  \bibinfo{booktitle}{\emph{Proceedings of the 2021 ACM Conference on Fairness,
  Accountability, and Transparency}}. \bibinfo{pages}{598--609}.
\newblock


\bibitem[Cobbe and Singh(2021)]%
        {cobbe2021artificial}
\bibfield{author}{\bibinfo{person}{Jennifer Cobbe} {and}
  \bibinfo{person}{Jatinder Singh}.} \bibinfo{year}{2021}\natexlab{}.
\newblock \showarticletitle{Artificial Intelligence as a Service: Legal
  Responsibilities, Liabilities, and Policy Challenges}.
\newblock \bibinfo{journal}{\emph{Forthcoming in Computer Law \& Security
  Review}} (\bibinfo{year}{2021}).
\newblock


\bibitem[Cobbe and Singh(2023)]%
        {cobbeDPDW}
\bibfield{author}{\bibinfo{person}{Jennifer Cobbe} {and}
  \bibinfo{person}{Jatinder Singh}.} \bibinfo{year}{2023}\natexlab{}.
\newblock \showarticletitle{Data Protection Doesn't Work}.
\newblock  (\bibinfo{year}{2023}).
\newblock
\urldef\tempurl%
\url{https://papers.ssrn.com/sol3/papers.cfm?abstract_id=4437133}
\showURL{%
\tempurl}


\bibitem[Cohen(2012)]%
        {cohenPiracySecurityArchitectures2012}
\bibfield{author}{\bibinfo{person}{Julie~E Cohen}.}
  \bibinfo{year}{2012}\natexlab{}.
\newblock \showarticletitle{``{{Piracy}},'' ``{{Security}},'' and
  {{Architectures}} of {{Control}}}.
\newblock In \bibinfo{booktitle}{\emph{Configuring the {{Networked Self}}:
  {{Law}}, {{Code}}, and the {{Play}} of {{Everyday Practice}}}}.
  \bibinfo{publisher}{{Yale University Press}}, \bibinfo{address}{{New Haven,
  CT}}.
\newblock


\bibitem[Cohen(2019)]%
        {cohen2019}
\bibfield{author}{\bibinfo{person}{Julie~E Cohen}.}
  \bibinfo{year}{2019}\natexlab{}.
\newblock \bibinfo{booktitle}{\emph{{Between Truth and Power: The Legal
  Constructions of Informational Capitalism}}}.
\newblock \bibinfo{publisher}{{Oxford University Press}}.
\newblock


\bibitem[Commission et~al\mbox{.}(2021)]%
        {EU-AIAct}
\bibfield{author}{\bibinfo{person}{EU Commission} {et~al\mbox{.}}}
  \bibinfo{year}{2021}\natexlab{}.
\newblock \showarticletitle{Proposal for a regulation of the European
  Parliament and of the Council laying down harmonised rules on artificial
  intelligence (Artificial Intelligence Act) and amending certain Union
  legislative acts}.
\newblock \bibinfo{journal}{\emph{COM (2021)}}  \bibinfo{volume}{206}
  (\bibinfo{year}{2021}).
\newblock


\bibitem[Crisan et~al\mbox{.}(2022)]%
        {crisan2022interactivemc}
\bibfield{author}{\bibinfo{person}{Anamaria Crisan}, \bibinfo{person}{Margaret
  Drouhard}, \bibinfo{person}{Jesse Vig}, {and} \bibinfo{person}{Nazneen
  Rajani}.} \bibinfo{year}{2022}\natexlab{}.
\newblock \showarticletitle{Interactive Model Cards: A Human-Centered Approach
  to Model Documentation}. In \bibinfo{booktitle}{\emph{2022 ACM Conference on
  Fairness, Accountability, and Transparency}} (Seoul, Republic of Korea)
  \emph{(\bibinfo{series}{FAccT '22})}. \bibinfo{publisher}{Association for
  Computing Machinery}, \bibinfo{address}{New York, NY, USA},
  \bibinfo{pages}{427–439}.
\newblock
\showISBNx{9781450393522}
\urldef\tempurl%
\url{https://doi.org/10.1145/3531146.3533108}
\showDOI{\tempurl}


\bibitem[Davis(2020)]%
        {davis2020}
\bibfield{author}{\bibinfo{person}{Jenny Davis}.}
  \bibinfo{year}{2020}\natexlab{}.
\newblock \bibinfo{booktitle}{\emph{{How Artifacts Afford: The Power and
  Politics of Everyday Things}}}.
\newblock \bibinfo{publisher}{{MIT Press}}.
\newblock


\bibitem[{Demos et al}(2020)]%
        {demos2020}
\bibfield{author}{\bibinfo{person}{{Demos et al}}.}
  \bibinfo{year}{2020}\natexlab{}.
\newblock \bibinfo{title}{Algorithm Inspection and Regulatory Access}.
\newblock
\newblock


\bibitem[Diver(2018)]%
        {diver2018}
\bibfield{author}{\bibinfo{person}{Laurence Diver}.}
  \bibinfo{year}{2018}\natexlab{}.
\newblock \showarticletitle{{Law as a User: Design, Affordance, and the
  Technological Mediation of Norms}}.
\newblock \bibinfo{journal}{\emph{{SCRIPTed: A Journal of Law, Technology \&
  Society }}} (\bibinfo{year}{2018}).
\newblock
Issue 1.


\bibitem[{EAIDB:The Ethical AI Database }(2023)]%
        {ethicaldb}
\bibfield{author}{\bibinfo{person}{{EAIDB: The Ethical AI Database }}.}
  \bibinfo{year}{2023}\natexlab{}.
\newblock \bibinfo{title}{Market Map}.
\newblock \bibinfo{howpublished}
https://www.eaidb.org/map.html.
\newblock


\bibitem[Edwards and Veale(2017)]%
        {edwards2017slave}
\bibfield{author}{\bibinfo{person}{Lilian Edwards} {and}
  \bibinfo{person}{Michael Veale}.} \bibinfo{year}{2017}\natexlab{}.
\newblock \showarticletitle{Slave to the algorithm: Why a right to an
  explanation is probably not the remedy you are looking for}.
\newblock \bibinfo{journal}{\emph{Duke L. \& Tech. Rev.}}  \bibinfo{volume}{16}
  (\bibinfo{year}{2017}), \bibinfo{pages}{18}.
\newblock


\bibitem[Elias(1984)]%
        {elias1984}
\bibfield{author}{\bibinfo{person}{Norbert Elias}.}
  \bibinfo{year}{1984}\natexlab{}.
\newblock \bibinfo{booktitle}{\emph{What Is Sociology?}
  (\bibinfo{edition}{revised} ed.)}.
\newblock \bibinfo{publisher}{Columbia University Press}.
\newblock


\bibitem[{European Union}(2016)]%
        {EU-GDPR}
\bibfield{author}{\bibinfo{person}{{European Union}}.}
  \bibinfo{year}{2016}\natexlab{}.
\newblock \showarticletitle{{Regulation (EU) 2016/679 of the European
  Parliament and of the Council of 27 April 2016 on the protection of natural
  persons with regard to the processing of personal data and on the free
  movement of such data, and repealing Directive 95/46/EC (General Data
  Protection Regulation)}}.
\newblock \bibinfo{journal}{\emph{OJEU}}  \bibinfo{volume}{L119}
  (\bibinfo{date}{4 May} \bibinfo{year}{2016}), \bibinfo{pages}{1--88}.
\newblock


\bibitem[Face(2022)]%
        {HuggingFace}
\bibfield{author}{\bibinfo{person}{Hugging Face}.}
  \bibinfo{year}{2022}\natexlab{}.
\newblock \bibinfo{title}{The AI community building the future.}
\newblock \bibinfo{howpublished}{\url{https://huggingface.co}}.
\newblock


\bibitem[Gandy(2010)]%
        {gandyEngagingRationalDiscrimination2010}
\bibfield{author}{\bibinfo{person}{Oscar~H. Gandy}.}
  \bibinfo{year}{2010}\natexlab{}.
\newblock \showarticletitle{Engaging Rational Discrimination: Exploring Reasons
  for Placing Regulatory Constraints on Decision Support Systems}.
\newblock \bibinfo{journal}{\emph{Ethics and Information Technology}}
  \bibinfo{volume}{12}, \bibinfo{number}{1} (\bibinfo{date}{March}
  \bibinfo{year}{2010}), \bibinfo{pages}{29--42}.
\newblock
\urldef\tempurl%
\url{https://doi.org/10/bzwqrx}
\showDOI{\tempurl}


\bibitem[Gebru et~al\mbox{.}(2018)]%
        {gebru2018datasheets}
\bibfield{author}{\bibinfo{person}{Timnit Gebru}, \bibinfo{person}{Jamie
  Morgenstern}, \bibinfo{person}{Briana Vecchione},
  \bibinfo{person}{Jennifer~Wortman Vaughan}, \bibinfo{person}{Hanna Wallach},
  \bibinfo{person}{Hal Daum{\'e}~III}, {and} \bibinfo{person}{Kate Crawford}.}
  \bibinfo{year}{2018}\natexlab{}.
\newblock \showarticletitle{Datasheets for datasets}.
\newblock \bibinfo{journal}{\emph{arXiv preprint arXiv:1803.09010}}
  (\bibinfo{year}{2018}).
\newblock


\bibitem[Gibbs(2014)]%
        {guardiandeepmind}
\bibfield{author}{\bibinfo{person}{Samuel Gibbs}.}
  \bibinfo{year}{2014}\natexlab{}.
\newblock \bibinfo{title}{{Google buys UK artificial intelligence startup
  Deepmind for £400m}}.
\newblock
  \bibinfo{howpublished}{https://www.theguardian.com/technology/2014/jan/27/google-acquires-uk-artificial-intelligence-startup-deepmind}.
\newblock


\bibitem[Gibson(2014)]%
        {gibson1979}
\bibfield{author}{\bibinfo{person}{James~J Gibson}.}
  \bibinfo{year}{{2014}}\natexlab{}.
\newblock \bibinfo{booktitle}{\emph{{The Ecological Approach to Visual
  Perception}} (\bibinfo{edition}{{Classic}} ed.)}.
\newblock \bibinfo{publisher}{{Taylor Francis}}.
\newblock


\bibitem[{Google}(2021)]%
        {GoogleCloudAI}
\bibfield{author}{\bibinfo{person}{{Google}}.} \bibinfo{year}{2021}\natexlab{}.
\newblock \bibinfo{title}{Cloud AI Building Blocks}.
\newblock
  \bibinfo{howpublished}{https://cloud.google.com/products/ai/building-blocks}.
\newblock
\newblock
\shownote{(Accessed on 04/11/2021)}.


\bibitem[{Government of Canada}(2020)]%
        {canadaAlgAssess}
\bibfield{author}{\bibinfo{person}{{Government of Canada}}.}
  \bibinfo{year}{2020}\natexlab{}.
\newblock \bibinfo{title}{Algorithmic Impact Assessment}.
\newblock
\newblock


\bibitem[Gray and Suri(2019)]%
        {grayGhostWorkHow2019}
\bibfield{author}{\bibinfo{person}{Mary~L Gray} {and}
  \bibinfo{person}{Siddharth Suri}.} \bibinfo{year}{2019}\natexlab{}.
\newblock \bibinfo{booktitle}{\emph{Ghost {{Work}}: {{How}} to {{Stop Silicon
  Valley}} from {{Building}} a {{New Global Underclass}}}}.
\newblock \bibinfo{publisher}{{Houghton Mifflin Harcourt}},
  \bibinfo{address}{{Boston}}.
\newblock


\bibitem[Gürses and van Hoboken(2017)]%
        {gurses2017}
\bibfield{author}{\bibinfo{person}{Seda Gürses} {and} \bibinfo{person}{Joris
  van Hoboken}.} \bibinfo{year}{2017}\natexlab{}.
\newblock \showarticletitle{Privacy After the Agile Turn}.
\newblock In \bibinfo{booktitle}{\emph{Cambridge Handbook of Consumer
  Privacy}}, \bibfield{editor}{\bibinfo{person}{Jules Polonetsky},
  \bibinfo{person}{Omer Tene}, {and} \bibinfo{person}{Evan Selinger}} (Eds.).
\newblock


\bibitem[Hale(2019)]%
        {forbes2019}
\bibfield{author}{\bibinfo{person}{Kori Hale}.}
  \bibinfo{year}{2019}\natexlab{}.
\newblock \bibinfo{title}{{Google \& Microsoft Banking On Africa's AI Labeling
  Workforce}}.
\newblock
  \bibinfo{howpublished}{https://www.forbes.com/sites/korihale/2019/05/28/google-microsoft-banking-on-africas-ai-labeling-workforce}.
\newblock


\bibitem[Harcourt(2007)]%
        {harcourtPrediction}
\bibfield{author}{\bibinfo{person}{Bernard~E Harcourt}.}
  \bibinfo{year}{2007}\natexlab{}.
\newblock \bibinfo{booktitle}{\emph{Against {{Prediction}}}}.
\newblock \bibinfo{publisher}{{University of Chicago Press}},
  \bibinfo{address}{{Chicago}}.
\newblock


\bibitem[{HireVue}(2022)]%
        {HireVue}
\bibfield{author}{\bibinfo{person}{{HireVue}}.}
  \bibinfo{year}{2022}\natexlab{}.
\newblock \bibinfo{title}{End-to-End Hiring Experience Platform: Video
  Interviewing, Conversational AI \& More | HireVue}.
\newblock \bibinfo{howpublished}{https://www.hirevue.com/}.
\newblock


\bibitem[Hutchinson et~al\mbox{.}(2021)]%
        {hutchinson2021towards}
\bibfield{author}{\bibinfo{person}{Ben Hutchinson}, \bibinfo{person}{Andrew
  Smart}, \bibinfo{person}{Alex Hanna}, \bibinfo{person}{Emily Denton},
  \bibinfo{person}{Christina Greer}, \bibinfo{person}{Oddur Kjartansson},
  \bibinfo{person}{Parker Barnes}, {and} \bibinfo{person}{Margaret Mitchell}.}
  \bibinfo{year}{2021}\natexlab{}.
\newblock \showarticletitle{Towards accountability for machine learning
  datasets: Practices from software engineering and infrastructure}. In
  \bibinfo{booktitle}{\emph{Proceedings of the 2021 ACM Conference on Fairness,
  Accountability, and Transparency}}. \bibinfo{pages}{560--575}.
\newblock


\bibitem[{IBM}(2021)]%
        {IBMWatsonAI}
\bibfield{author}{\bibinfo{person}{{IBM}}.} \bibinfo{year}{2021}\natexlab{}.
\newblock \bibinfo{title}{IBM Watson products and solutions}.
\newblock
  \bibinfo{howpublished}{https://www.ibm.com/uk-en/watson/products-services}.
\newblock


\bibitem[Infermedica(2022)]%
        {Infermedica}
\bibfield{author}{\bibinfo{person}{Infermedica}.}
  \bibinfo{year}{2022}\natexlab{}.
\newblock \bibinfo{title}{Call Center Triage}.
\newblock
  \bibinfo{howpublished}{https://infermedica.com/product/call-center-triage}.
\newblock


\bibitem[{Information Commissioner's Office and The Alan Turing
  Institute}(2020)]%
        {ico2020}
\bibfield{author}{\bibinfo{person}{{Information Commissioner's Office and The
  Alan Turing Institute}}.} \bibinfo{year}{2020}\natexlab{}.
\newblock \bibinfo{title}{{Explaining decisions made with AI}}.
\newblock
\newblock


\bibitem[Javadi et~al\mbox{.}(2020)]%
        {javadi2020monitoring}
\bibfield{author}{\bibinfo{person}{Seyyed~Ahmad Javadi},
  \bibinfo{person}{Richard Cloete}, \bibinfo{person}{Jennifer Cobbe},
  \bibinfo{person}{Michelle Seng~Ah Lee}, {and} \bibinfo{person}{Jatinder
  Singh}.} \bibinfo{year}{2020}\natexlab{}.
\newblock \bibinfo{booktitle}{\emph{Monitoring Misuse for Accountable
  'Artificial Intelligence as a Service'}}.
\newblock \bibinfo{publisher}{Association for Computing Machinery},
  \bibinfo{address}{New York, NY, USA}, \bibinfo{pages}{300–306}.
\newblock
\showISBNx{9781450371100}
\urldef\tempurl%
\url{https://doi.org/10.1145/3375627.3375873}
\showURL{%
\tempurl}


\bibitem[Javadi et~al\mbox{.}(2021)]%
        {javadi2021monitoring}
\bibfield{author}{\bibinfo{person}{Seyyed~Ahmad Javadi}, \bibinfo{person}{Chris
  Norval}, \bibinfo{person}{Richard Cloete}, {and} \bibinfo{person}{Jatinder
  Singh}.} \bibinfo{year}{2021}\natexlab{}.
\newblock \showarticletitle{Monitoring AI Services for Misuse}. In
  \bibinfo{booktitle}{\emph{Proceedings of the 2021 AAAI/ACM Conference on AI,
  Ethics, and Society}}. \bibinfo{pages}{597--607}.
\newblock


\bibitem[Kamarinou et~al\mbox{.}(2015)]%
        {kamarinou2015}
\bibfield{author}{\bibinfo{person}{Dimitra Kamarinou},
  \bibinfo{person}{Christopher Millard}, {and} \bibinfo{person}{Kuan Hon, W}.}
  \bibinfo{year}{2015}\natexlab{}.
\newblock \showarticletitle{Privacy in the Clouds: An Empirical Study of the
  Terms of Service and Privacy Policies of 20 Cloud Service Providers}.
\newblock \bibinfo{journal}{\emph{{Queen Mary School of Law Legal Studies
  Research Paper No. 209/2015}}} (\bibinfo{year}{2015}).
\newblock
\urldef\tempurl%
\url{https://papers.ssrn.com/sol3/papers.cfm?abstract_id=2646447}
\showURL{%
\tempurl}


\bibitem[Kaminski and Urban(2021)]%
        {kaminski2021}
\bibfield{author}{\bibinfo{person}{Margot Kaminski} {and}
  \bibinfo{person}{Jennifer~M Urban}.} \bibinfo{year}{2021}\natexlab{}.
\newblock \showarticletitle{The Right to Contest AI}.
\newblock \bibinfo{journal}{\emph{Columbia Law Review}} \bibinfo{volume}{121},
  \bibinfo{number}{7} (\bibinfo{year}{2021}).
\newblock


\bibitem[Koene et~al\mbox{.}(2019)]%
        {koene2019}
\bibfield{author}{\bibinfo{person}{Ansgar Koene} {et~al\mbox{.}}}
  \bibinfo{year}{2019}\natexlab{}.
\newblock \showarticletitle{A governance framework for algorithmic
  accountability and transparency}.
\newblock \bibinfo{journal}{\emph{European Parliamentary Research Service,
  Panel for the Future of Science and Technology}}  \bibinfo{volume}{PE
  624.262} (\bibinfo{date}{April} \bibinfo{year}{2019}).
\newblock


\bibitem[Kostova et~al\mbox{.}(2020)]%
        {kostovaPrivacyEngineeringMeets2020a}
\bibfield{author}{\bibinfo{person}{Blagovesta Kostova}, \bibinfo{person}{Seda
  G{\"u}rses}, {and} \bibinfo{person}{Carmela Troncoso}.}
  \bibinfo{year}{2020}\natexlab{}.
\newblock \bibinfo{title}{Privacy {{Engineering Meets Software Engineering}}.
  {{On}} the {{Challenges}} of {{Engineering Privacy ByDesign}}}.
\newblock
\newblock
\urldef\tempurl%
\url{https://doi.org/10.48550/arXiv.2007.08613}
\showDOI{\tempurl}
\showeprint[arxiv]{2007.08613}~[cs]


\bibitem[Krafft et~al\mbox{.}(2021)]%
        {krafft2021action}
\bibfield{author}{\bibinfo{person}{PM Krafft}, \bibinfo{person}{Meg Young},
  \bibinfo{person}{Michael Katell}, \bibinfo{person}{Jennifer~E Lee},
  \bibinfo{person}{Shankar Narayan}, \bibinfo{person}{Micah Epstein},
  \bibinfo{person}{Dharma Dailey}, \bibinfo{person}{Bernease Herman},
  \bibinfo{person}{Aaron Tam}, \bibinfo{person}{Vivian Guetler},
  {et~al\mbox{.}}} \bibinfo{year}{2021}\natexlab{}.
\newblock \showarticletitle{An Action-Oriented AI Policy Toolkit for Technology
  Audits by Community Advocates and Activists}. In
  \bibinfo{booktitle}{\emph{Proceedings of the 2021 ACM Conference on Fairness,
  Accountability, and Transparency}}. \bibinfo{pages}{772--781}.
\newblock


\bibitem[Kroll et~al\mbox{.}(2017)]%
        {krollaa}
\bibfield{author}{\bibinfo{person}{{Joshua A.} Kroll}, \bibinfo{person}{Joanna
  Huey}, \bibinfo{person}{Solon Barocas}, \bibinfo{person}{{Edward W.} Felten},
  \bibinfo{person}{{Joel R.} Reidenberg}, \bibinfo{person}{{David G.}
  Robinson}, {and} \bibinfo{person}{Harlan Yu}.}
  \bibinfo{year}{2017}\natexlab{}.
\newblock \showarticletitle{Accountable algorithms}.
\newblock \bibinfo{journal}{\emph{University of Pennsylvania Law Review}}
  \bibinfo{volume}{165}, \bibinfo{number}{3} (\bibinfo{date}{Feb.}
  \bibinfo{year}{2017}), \bibinfo{pages}{633--705}.
\newblock
\showISSN{0041-9907}


\bibitem[Kroll(2020)]%
        {krollAccountabilityComputerSystems2020}
\bibfield{author}{\bibinfo{person}{Joshua~A. Kroll}.}
  \bibinfo{year}{2020}\natexlab{}.
\newblock \showarticletitle{Accountability in {{Computer Systems}}}.
\newblock In \bibinfo{booktitle}{\emph{The {{Oxford Handbook}} of {{Ethics}} of
  {{AI}}}}, \bibfield{editor}{\bibinfo{person}{Markus~D. Dubber},
  \bibinfo{person}{Frank Pasquale}, {and} \bibinfo{person}{Sunit Das}} (Eds.).
  \bibinfo{publisher}{{Oxford University Press}}, \bibinfo{pages}{0}.
\newblock
\showISBNx{978-0-19-006739-7}
\urldef\tempurl%
\url{https://doi.org/10.1093/oxfordhb/9780190067397.013.10}
\showDOI{\tempurl}


\bibitem[Kroll(2021)]%
        {kroll2021traceability}
\bibfield{author}{\bibinfo{person}{Joshua~A. Kroll}.}
  \bibinfo{year}{2021}\natexlab{}.
\newblock \showarticletitle{Outlining Traceability: A Principle for
  Operationalizing Accountability in Computing Systems}. In
  \bibinfo{booktitle}{\emph{Proceedings of the 2021 ACM Conference on Fairness,
  Accountability, and Transparency}} (Virtual Event, Canada)
  \emph{(\bibinfo{series}{FAccT '21})}. \bibinfo{publisher}{Association for
  Computing Machinery}, \bibinfo{address}{New York, NY, USA},
  \bibinfo{pages}{758–771}.
\newblock
\showISBNx{9781450383097}
\urldef\tempurl%
\url{https://doi.org/10.1145/3442188.3445937}
\showDOI{\tempurl}


\bibitem[Lee and Singh(2021)]%
        {lee2021risk}
\bibfield{author}{\bibinfo{person}{Michelle Seng~Ah Lee} {and}
  \bibinfo{person}{Jatinder Singh}.} \bibinfo{year}{2021}\natexlab{}.
\newblock \showarticletitle{Risk identification questionnaire for unintended
  bias in machine learning development lifecycle}.
\newblock \bibinfo{journal}{\emph{Available at SSRN}} (\bibinfo{year}{2021}).
\newblock


\bibitem[Lewicki et~al\mbox{.}(2023)]%
        {lewicki2023aiaasfairness}
\bibfield{author}{\bibinfo{person}{Kornel Lewicki}, \bibinfo{person}{Michelle
  Seng Ah~Lee}, \bibinfo{person}{Jennifer Cobbe}, {and}
  \bibinfo{person}{Jatinder Singh}.} \bibinfo{year}{{2023}}\natexlab{}.
\newblock \showarticletitle{{Out of Context: Algorithmic Fairness in
  "Artificial Intelligence as a Service"}}.
\newblock \bibinfo{journal}{\emph{{arXiv:2302.01448}}}
  (\bibinfo{year}{{2023}}).
\newblock
https://arxiv.org/abs/2302.01448


\bibitem[Lunit(2022)]%
        {Lunit75:online}
\bibfield{author}{\bibinfo{person}{Lunit}.} \bibinfo{year}{2022}\natexlab{}.
\newblock \bibinfo{title}{AI will be the new standard of care. By Lunit.}
\newblock \bibinfo{howpublished}{\url{https://www.lunit.io/en}}.
\newblock


\bibitem[Lyon(2003)]%
        {lyonSurveillanceSocialSorting2003}
\bibfield{editor}{\bibinfo{person}{David Lyon}} (Ed.).
  \bibinfo{year}{2003}\natexlab{}.
\newblock \bibinfo{booktitle}{\emph{Surveillance as Social Sorting: Privacy,
  Risk, and Digital Discrimination}}.
\newblock \bibinfo{publisher}{{Routledge}}, \bibinfo{address}{{London ; New
  York}}.
\newblock
\showISBNx{978-0-415-27872-0 978-0-415-27873-7}
\showLCCN{JC596 .S796 2003}


\bibitem[Madaio et~al\mbox{.}(2020)]%
        {madaio2020co}
\bibfield{author}{\bibinfo{person}{Michael~A Madaio}, \bibinfo{person}{Luke
  Stark}, \bibinfo{person}{Jennifer Wortman~Vaughan}, {and}
  \bibinfo{person}{Hanna Wallach}.} \bibinfo{year}{2020}\natexlab{}.
\newblock \showarticletitle{Co-designing checklists to understand
  organizational challenges and opportunities around fairness in ai}. In
  \bibinfo{booktitle}{\emph{Proceedings of the 2020 CHI Conference on Human
  Factors in Computing Systems}}. \bibinfo{pages}{1--14}.
\newblock


\bibitem[Mahieu et~al\mbox{.}(2019)]%
        {mahieu2019}
\bibfield{author}{\bibinfo{person}{René Mahieu}, \bibinfo{person}{Joris van
  Hoboken}, {and} \bibinfo{person}{Hadi Asghari}.}
  \bibinfo{year}{2019}\natexlab{}.
\newblock \showarticletitle{{Responsibility for Data Protection in a Networked
  World: On the Queston of the Controller, “Effective and Complete
  Protection” and its Application to Data Access Rights in Europe}}.
\newblock \bibinfo{journal}{\emph{{Journal of Intellectual Property,
  Information Technology and E-Commerce Law}}}  \bibinfo{volume}{10}
  (\bibinfo{year}{2019}).
\newblock
Issue 1.


\bibitem[Matus and Veale(2022)]%
        {matusCertificationSystemsMachine2022}
\bibfield{author}{\bibinfo{person}{Kira J.~M. Matus} {and}
  \bibinfo{person}{Michael Veale}.} \bibinfo{year}{2022}\natexlab{}.
\newblock \showarticletitle{Certification Systems for Machine Learning:
  {{Lessons}} from Sustainability}.
\newblock \bibinfo{journal}{\emph{Regulation \& Governance}}
  \bibinfo{volume}{16}, \bibinfo{number}{1} (\bibinfo{year}{2022}),
  \bibinfo{pages}{177--196}.
\newblock
\urldef\tempurl%
\url{https://doi.org/10.1111/rego.12417}
\showDOI{\tempurl}


\bibitem[Metaxa et~al\mbox{.}(2021)]%
        {metaxa2021auditing}
\bibfield{author}{\bibinfo{person}{Dana{\"e} Metaxa},
  \bibinfo{person}{Joon~Sung Park}, \bibinfo{person}{Ronald~E Robertson},
  \bibinfo{person}{Karrie Karahalios}, \bibinfo{person}{Christo Wilson},
  \bibinfo{person}{Jeff Hancock}, \bibinfo{person}{Christian Sandvig},
  {et~al\mbox{.}}} \bibinfo{year}{2021}\natexlab{}.
\newblock \showarticletitle{Auditing algorithms: Understanding algorithmic
  systems from the outside in}.
\newblock \bibinfo{journal}{\emph{Foundations and Trends{\textregistered} in
  Human--Computer Interaction}} \bibinfo{volume}{14}, \bibinfo{number}{4}
  (\bibinfo{year}{2021}), \bibinfo{pages}{272--344}.
\newblock


\bibitem[Michels et~al\mbox{.}(2021)]%
        {michelsStandardContractsCloud2021}
\bibfield{author}{\bibinfo{person}{Johan~David Michels},
  \bibinfo{person}{Christopher Millard}, {and} \bibinfo{person}{Felicity
  Turton}.} \bibinfo{year}{2021}\natexlab{}.
\newblock \showarticletitle{Standard {{Contracts}} for {{Cloud Services}}}.
\newblock In \bibinfo{booktitle}{\emph{Cloud {{Computing Law}}}
  (\bibinfo{edition}{second} ed.)},
  \bibfield{editor}{\bibinfo{person}{Christopher Millard}} (Ed.).
  \bibinfo{publisher}{{Oxford University Press}}, \bibinfo{address}{{Oxford}},
  \bibinfo{pages}{49--99}.
\newblock
\showISBNx{978-0-19-871666-2 978-0-19-191858-2}
\urldef\tempurl%
\url{https://doi.org/10.1093/oso/9780198716662.003.0003}
\showDOI{\tempurl}


\bibitem[{Microsoft}(2021)]%
        {MicrosoftCognitive}
\bibfield{author}{\bibinfo{person}{{Microsoft}}.}
  \bibinfo{year}{2021}\natexlab{}.
\newblock \bibinfo{title}{Cognitive Services}.
\newblock
  \bibinfo{howpublished}{\url{https://azure.microsoft.com/en-gb/services/cognitive-services}}.
\newblock
\newblock
\shownote{(Accessed on 04/11/2021)}.


\bibitem[Microsoft(2023)]%
        {msopenai}
\bibfield{author}{\bibinfo{person}{Microsoft}.}
  \bibinfo{year}{2023}\natexlab{}.
\newblock \bibinfo{title}{{Microsoft and OpenAI extend partnership}}.
\newblock
  \bibinfo{howpublished}{\url{https://blogs.microsoft.com/blog/2023/01/23/microsoftandopenaiextendpartnership}}.
\newblock


\bibitem[Millard(2021)]%
        {millard2021}
\bibfield{author}{\bibinfo{person}{Christopher Millard}.}
  \bibinfo{year}{2021}\natexlab{}.
\newblock \bibinfo{booktitle}{\emph{Cloud Computing Law} (\bibinfo{edition}{2}
  ed.)}.
\newblock \bibinfo{publisher}{Oxford University Press}.
\newblock


\bibitem[Mitchell et~al\mbox{.}(2019)]%
        {mitchell2019model}
\bibfield{author}{\bibinfo{person}{Margaret Mitchell}, \bibinfo{person}{Simone
  Wu}, \bibinfo{person}{Andrew Zaldivar}, \bibinfo{person}{Parker Barnes},
  \bibinfo{person}{Lucy Vasserman}, \bibinfo{person}{Ben Hutchinson},
  \bibinfo{person}{Elena Spitzer}, \bibinfo{person}{Inioluwa~Deborah Raji},
  {and} \bibinfo{person}{Timnit Gebru}.} \bibinfo{year}{2019}\natexlab{}.
\newblock \showarticletitle{Model cards for model reporting}. In
  \bibinfo{booktitle}{\emph{Conference on Fairness, Accountability, and
  Transparency}}. \bibinfo{pages}{220--229}.
\newblock


\bibitem[Mosco(2014)]%
        {moscoDarkClouds2014}
\bibfield{author}{\bibinfo{person}{Vincent Mosco}.}
  \bibinfo{year}{2014}\natexlab{}.
\newblock \showarticletitle{Dark {{Clouds}}}.
\newblock In \bibinfo{booktitle}{\emph{To the {{Cloud}}: {{Big Data}} in a
  {{Turbulent World}}}}. \bibinfo{publisher}{{Paradigm}},
  \bibinfo{address}{{Boulder, CO}}.
\newblock


\bibitem[Nissenbaum(1996)]%
        {nissenbaum1996}
\bibfield{author}{\bibinfo{person}{Helen Nissenbaum}.}
  \bibinfo{year}{1996}\natexlab{}.
\newblock \showarticletitle{Accountability in a Computerized Society}.
\newblock \bibinfo{journal}{\emph{Science and Engineering Ethics}}
  (\bibinfo{year}{1996}).
\newblock


\bibitem[Norman(1988)]%
        {norman1988}
\bibfield{author}{\bibinfo{person}{Donald Norman}.}
  \bibinfo{year}{1988}\natexlab{}.
\newblock \bibinfo{booktitle}{\emph{{The Design of Everyday Things}}}.
\newblock \bibinfo{publisher}{{Basic Books}}.
\newblock


\bibitem[Norval et~al\mbox{.}(2020)]%
        {norval2020reviewableIoT}
\bibfield{author}{\bibinfo{person}{Chris Norval}, \bibinfo{person}{Jennifer
  Cobbe}, {and} \bibinfo{person}{Jatinder Singh}.}
  \bibinfo{year}{2020}\natexlab{}.
\newblock \showarticletitle{Towards an accountable Internet of Things: A call
  for `reviewability'}. In \bibinfo{booktitle}{\emph{Privacy by Design for the
  Internet of Things: Building Accountability and Security}}.
  \bibinfo{publisher}{The Institution of Engineering and Technology}.
\newblock


\bibitem[Norval et~al\mbox{.}(2022)]%
        {norval2022disclosures}
\bibfield{author}{\bibinfo{person}{Chris Norval}, \bibinfo{person}{Kristin
  Cornelius}, \bibinfo{person}{Jennifer Cobbe}, {and} \bibinfo{person}{Jatinder
  Singh}.} \bibinfo{year}{2022}\natexlab{}.
\newblock \showarticletitle{Disclosure by Design: Designing Information
  Disclosures to Support Meaningful Transparency and Accountability}. In
  \bibinfo{booktitle}{\emph{2022 ACM Conference on Fairness, Accountability,
  and Transparency}} (Seoul, Republic of Korea) \emph{(\bibinfo{series}{FAccT
  '22})}. \bibinfo{publisher}{Association for Computing Machinery},
  \bibinfo{address}{New York, NY, USA}, \bibinfo{pages}{679–690}.
\newblock
\showISBNx{9781450393522}
\urldef\tempurl%
\url{https://doi.org/10.1145/3531146.3533133}
\showDOI{\tempurl}


\bibitem[Pasquale(2019)]%
        {pasquale2019}
\bibfield{author}{\bibinfo{person}{Frank Pasquale}.}
  \bibinfo{year}{2019}\natexlab{}.
\newblock \showarticletitle{The Second Wave of Algorithmic Accountability}.
\newblock \bibinfo{journal}{\emph{Law and Political Economy Project}}
  (\bibinfo{year}{2019}).
\newblock
\urldef\tempurl%
\url{https://lpeproject.org/blog/the-second-wave-of-algorithmic-accountability/}
\showURL{%
\tempurl}


\bibitem[Passi and Barocas(2019)]%
        {passiProblemFormulationFairness2019}
\bibfield{author}{\bibinfo{person}{Samir Passi} {and} \bibinfo{person}{Solon
  Barocas}.} \bibinfo{year}{2019}\natexlab{}.
\newblock \showarticletitle{Problem {{Formulation}} and {{Fairness}}}. In
  \bibinfo{booktitle}{\emph{Proceedings of the {{Conference}} on {{Fairness}},
  {{Accountability}}, and {{Transparency}}}} \emph{(\bibinfo{series}{{{FAT}}*
  '19})}. \bibinfo{publisher}{{ACM}}, \bibinfo{address}{{New York, NY, USA}},
  \bibinfo{pages}{39--48}.
\newblock
\showISBNx{978-1-4503-6125-5}
\urldef\tempurl%
\url{https://doi.org/10/gftmpv}
\showDOI{\tempurl}


\bibitem[Perrigo(n.\,d.)]%
        {timeOpenAI}
\bibfield{author}{\bibinfo{person}{Billy Perrigo}.}
  \bibinfo{year}{[n.\,d.]}\natexlab{}.
\newblock \showarticletitle{OpenAI Used Kenyan Workers on Less Than \$2 Per
  Hour to Make ChatGPT Less Toxic}.
\newblock \bibinfo{journal}{\emph{TIME}} (\bibinfo{year}{[n.\,d.]}).
\newblock
\urldef\tempurl%
\url{https://time.com/6247678/openai-chatgpt-kenya-workers/}
\showURL{%
\tempurl}


\bibitem[Perrigo(2023)]%
        {time2023}
\bibfield{author}{\bibinfo{person}{Billy Perrigo}.}
  \bibinfo{year}{2023}\natexlab{}.
\newblock \bibinfo{title}{{Exclusive: OpenAI Used Kenyan Workers on Less Than
  \$2 Per Hour to Make ChatGPT Less Toxic}}.
\newblock
  \bibinfo{howpublished}{https://time.com/6247678/openai-chatgpt-kenya-workers}.
\newblock


\bibitem[Pushkarna et~al\mbox{.}(2022a)]%
        {pushkarna2019datacards}
\bibfield{author}{\bibinfo{person}{Mahima Pushkarna}, \bibinfo{person}{Andrew
  Zaldivar}, {and} \bibinfo{person}{Oddur Kjartansson}.}
  \bibinfo{year}{2022}\natexlab{a}.
\newblock \showarticletitle{Data Cards: Purposeful and Transparent Dataset
  Documentation for Responsible AI}. In \bibinfo{booktitle}{\emph{Conference on
  Fairness, Accountability, and Transparency}}.
\newblock


\bibitem[Pushkarna et~al\mbox{.}(2022b)]%
        {pushkarna2022datacards}
\bibfield{author}{\bibinfo{person}{Mahima Pushkarna}, \bibinfo{person}{Andrew
  Zaldivar}, {and} \bibinfo{person}{Oddur Kjartansson}.}
  \bibinfo{year}{2022}\natexlab{b}.
\newblock \showarticletitle{Data Cards: Purposeful and Transparent Dataset
  Documentation for Responsible AI} \emph{(\bibinfo{series}{FAccT '22})}.
  \bibinfo{publisher}{Association for Computing Machinery},
  \bibinfo{address}{New York, NY, USA}, \bibinfo{pages}{1776–1826}.
\newblock
\showISBNx{9781450393522}
\urldef\tempurl%
\url{https://doi.org/10.1145/3531146.3533231}
\showDOI{\tempurl}


\bibitem[{Pymetrics}(2021)]%
        {pymetrics}
\bibfield{author}{\bibinfo{person}{{Pymetrics}}.}
  \bibinfo{year}{2021}\natexlab{}.
\newblock \bibinfo{title}{Talent Matching Platform}.
\newblock \bibinfo{howpublished}{\url{https://www.pymetrics.ai/}}.
\newblock


\bibitem[Rader et~al\mbox{.}(2018)]%
        {rader2018explanations}
\bibfield{author}{\bibinfo{person}{Emilee Rader}, \bibinfo{person}{Kelley
  Cotter}, {and} \bibinfo{person}{Janghee Cho}.}
  \bibinfo{year}{2018}\natexlab{}.
\newblock \showarticletitle{Explanations as mechanisms for supporting
  algorithmic transparency}. In \bibinfo{booktitle}{\emph{Proceedings of the
  2018 CHI conference on human factors in computing systems}}.
  \bibinfo{pages}{1--13}.
\newblock


\bibitem[Raji et~al\mbox{.}(2020)]%
        {raji2020closing}
\bibfield{author}{\bibinfo{person}{Inioluwa~Deborah Raji},
  \bibinfo{person}{Andrew Smart}, \bibinfo{person}{Rebecca~N White},
  \bibinfo{person}{Margaret Mitchell}, \bibinfo{person}{Timnit Gebru},
  \bibinfo{person}{Ben Hutchinson}, \bibinfo{person}{Jamila Smith-Loud},
  \bibinfo{person}{Daniel Theron}, {and} \bibinfo{person}{Parker Barnes}.}
  \bibinfo{year}{2020}\natexlab{}.
\newblock \showarticletitle{Closing the AI accountability gap: defining an
  end-to-end framework for internal algorithmic auditing}. In
  \bibinfo{booktitle}{\emph{Proceedings of the 2020 Conference on Fairness,
  Accountability, and Transparency}}. \bibinfo{pages}{33--44}.
\newblock


\bibitem[Reed and Edgar(2021)]%
        {reedConsumerProtectionCloud2021}
\bibfield{author}{\bibinfo{person}{Chris Reed} {and} \bibinfo{person}{Laura
  Edgar}.} \bibinfo{year}{2021}\natexlab{}.
\newblock \showarticletitle{Consumer {{Protection}} in the {{Cloud}}}.
\newblock In \bibinfo{booktitle}{\emph{Cloud {{Computing Law}}}
  (\bibinfo{edition}{second} ed.)},
  \bibfield{editor}{\bibinfo{person}{Christopher Millard}} (Ed.).
  \bibinfo{publisher}{{Oxford University Press}}, \bibinfo{address}{{Oxford}},
  \bibinfo{pages}{218--254}.
\newblock
\showISBNx{978-0-19-871666-2 978-0-19-191858-2}
\urldef\tempurl%
\url{https://doi.org/10.1093/oso/9780198716662.003.0007}
\showDOI{\tempurl}


\bibitem[Reidenberg(1998)]%
        {reidenbergLexInformaticaFormulation1998}
\bibfield{author}{\bibinfo{person}{Joel~R. Reidenberg}.}
  \bibinfo{year}{1998}\natexlab{}.
\newblock \showarticletitle{Lex {{Informatica}}: {{The Formulation}} of
  {{Information Policy Rules}} through {{Technology}}}.
\newblock \bibinfo{journal}{\emph{Texas Law Review}} \bibinfo{volume}{76},
  \bibinfo{number}{3} (\bibinfo{year}{1998}), \bibinfo{pages}{553--594}.
\newblock


\bibitem[Reisman et~al\mbox{.}(2018)]%
        {ainow}
\bibfield{author}{\bibinfo{person}{Dillon Reisman} {et~al\mbox{.}}}
  \bibinfo{year}{2018}\natexlab{}.
\newblock \bibinfo{title}{Algorithmic Impact Assessments: A practical framework
  for public agency accountability (AI Now)}.
\newblock
\newblock


\bibitem[Richter(2022)]%
        {statista}
\bibfield{author}{\bibinfo{person}{Felix Richter}.}
  \bibinfo{year}{2022}\natexlab{}.
\newblock \bibinfo{title}{{Amazon, Microsoft \& Google Dominate Cloud Market}}.
\newblock
  \bibinfo{howpublished}{https://www.statista.com/chart/18819/worldwide-market-share-of-leading-cloud-infrastructure-service-providers}.
\newblock


\bibitem[Saleiro et~al\mbox{.}(2018)]%
        {saleiro2018aequitas}
\bibfield{author}{\bibinfo{person}{Pedro Saleiro}, \bibinfo{person}{Benedict
  Kuester}, \bibinfo{person}{Loren Hinkson}, \bibinfo{person}{Jesse London},
  \bibinfo{person}{Abby Stevens}, \bibinfo{person}{Ari Anisfeld},
  \bibinfo{person}{Kit~T Rodolfa}, {and} \bibinfo{person}{Rayid Ghani}.}
  \bibinfo{year}{2018}\natexlab{}.
\newblock \showarticletitle{Aequitas: A bias and fairness audit toolkit}.
\newblock \bibinfo{journal}{\emph{arXiv preprint arXiv:1811.05577}}
  (\bibinfo{year}{2018}).
\newblock


\bibitem[{Sama AI}(n.\,d.)]%
        {samaAI}
\bibfield{author}{\bibinfo{person}{{Sama AI}}.}
  \bibinfo{year}{[n.\,d.]}\natexlab{}.
\newblock \bibinfo{title}{{https://www.sama.com}}.
\newblock \bibinfo{howpublished}{{https://www.sama.com}}.
\newblock


\bibitem[Seaver(2013)]%
        {seaver2013}
\bibfield{author}{\bibinfo{person}{Nick Seaver}.}
  \bibinfo{year}{2013}\natexlab{}.
\newblock \showarticletitle{Knowing Algorithms}.
\newblock \bibinfo{journal}{\emph{paper presented at Media in Transition 8}}
  (\bibinfo{year}{2013}).
\newblock


\bibitem[Selbst et~al\mbox{.}(2019)]%
        {selbstFairnessAbstractionSociotechnical2019}
\bibfield{author}{\bibinfo{person}{Andrew~D. Selbst}, \bibinfo{person}{Danah
  Boyd}, \bibinfo{person}{Sorelle~A. Friedler}, \bibinfo{person}{Suresh
  Venkatasubramanian}, {and} \bibinfo{person}{Janet Vertesi}.}
  \bibinfo{year}{2019}\natexlab{}.
\newblock \showarticletitle{Fairness and {{Abstraction}} in {{Sociotechnical
  Systems}}}. In \bibinfo{booktitle}{\emph{Proceedings of the {{Conference}} on
  {{Fairness}}, {{Accountability}}, and {{Transparency}}}}
  \emph{(\bibinfo{series}{{{FAT}}* '19})}. \bibinfo{publisher}{{ACM}},
  \bibinfo{address}{{New York, NY, USA}}, \bibinfo{pages}{59--68}.
\newblock
\showISBNx{978-1-4503-6125-5}
\urldef\tempurl%
\url{https://doi.org/10/gftmp3}
\showDOI{\tempurl}


\bibitem[Singh et~al\mbox{.}(2016)]%
        {singh2016partofaprocess}
\bibfield{author}{\bibinfo{person}{Jatinder Singh} {et~al\mbox{.}}}
  \bibinfo{year}{2016}\natexlab{}.
\newblock \bibinfo{title}{Responsibility \& Machine Learning: Part of a
  Process}.
\newblock
\newblock


\bibitem[Singh et~al\mbox{.}(2019)]%
        {singh2019decprov}
\bibfield{author}{\bibinfo{person}{Jatinder Singh}, \bibinfo{person}{Jennifer
  Cobbe}, {and} \bibinfo{person}{Chris Norval}.}
  \bibinfo{year}{2019}\natexlab{}.
\newblock \showarticletitle{{Decision Provenance: Harnessing Data Flow for
  Accountable Systems}}.
\newblock \bibinfo{journal}{\emph{IEEE Access}}  \bibinfo{volume}{7}
  (\bibinfo{year}{2019}), \bibinfo{pages}{6562--6574}.
\newblock



\bibitem[Sivasubramanian(2023)]%
        {awsintermed}
\bibfield{author}{\bibinfo{person}{Swami Sivasubramanian}.}
  \bibinfo{year}{2023}\natexlab{}.
\newblock \bibinfo{title}{{Announcing New Tools for Building with Generative AI
  on AWS}}.
\newblock
  \bibinfo{howpublished}{https://aws.amazon.com/blogs/machine-learning/announcing-new-tools-for-building-with-generative-ai-on-aws/}.
\newblock


\bibitem[{State of Califorina}(2018)]%
        {US-CCPA}
\bibfield{author}{\bibinfo{person}{{State of Califorina}}.}
  \bibinfo{year}{2018}\natexlab{}.
\newblock \bibinfo{title}{{The California Consumer Privacy Act (CCPA)}}.
\newblock
\newblock


\bibitem[{Supahands}(n.\,d.)]%
        {supahands}
\bibfield{author}{\bibinfo{person}{{Supahands}}.}
  \bibinfo{year}{[n.\,d.]}\natexlab{}.
\newblock \bibinfo{title}{{https://www.supaagents.com}}.
\newblock \bibinfo{howpublished}{{https://www.supaagents.com}}.
\newblock


\bibitem[{The Economist}(2022)]%
        {ArtificialIntelligencePermeating2022}
\bibfield{author}{\bibinfo{person}{{The Economist}}.}
  \bibinfo{year}{2022}\natexlab{}.
\newblock \bibinfo{title}{Artificial Intelligence Is Permeating Business at
  Last}.
\newblock
\newblock
\urldef\tempurl%
\url{https://www.economist.com/business/2022/12/06/artificial-intelligence-is-permeating-business-at-last}
\showURL{%
\tempurl}


\bibitem[Thompson(1980)]%
        {thompson1980}
\bibfield{author}{\bibinfo{person}{Dennis~F. Thompson}.}
  \bibinfo{year}{1980}\natexlab{}.
\newblock \showarticletitle{Moral Responsibility of Public Officials: The
  Problem of Many Hands}.
\newblock \bibinfo{journal}{\emph{American Political Science Review}}
  \bibinfo{volume}{74}, \bibinfo{number}{4} (\bibinfo{year}{1980}),
  \bibinfo{pages}{905–916}.
\newblock
\urldef\tempurl%
\url{https://doi.org/10.2307/1954312}
\showDOI{\tempurl}


\bibitem[Turton et~al\mbox{.}(2021)]%
        {turton2021}
\bibfield{author}{\bibinfo{person}{Felicity Turton}, \bibinfo{person}{Dimitra
  Kamarinou}, \bibinfo{person}{Johan~David Michels}, {and}
  \bibinfo{person}{Christopher Millard}.} \bibinfo{year}{2021}\natexlab{}.
\newblock \showarticletitle{Privacy in the Clouds, Revisited: An Analysis of
  the Privacy Policies of 40 Cloud Computing Services}.
\newblock \bibinfo{journal}{\emph{{Queen Mary Law Research Paper No.
  354/2021}}} (\bibinfo{year}{2021}).
\newblock
\urldef\tempurl%
\url{https://papers.ssrn.com/sol3/papers.cfm?abstract_id=3823424}
\showURL{%
\tempurl}


\bibitem[Veale et~al\mbox{.}(2022)]%
        {vealeImpossibleAsksCan2022}
\bibfield{author}{\bibinfo{person}{Michael Veale}, \bibinfo{person}{Midas
  Nouwens}, {and} \bibinfo{person}{Cristiana Santos}.}
  \bibinfo{year}{2022}\natexlab{}.
\newblock \showarticletitle{Impossible {{Asks}}: {{Can}} the {{Transparency}}
  and {{Consent Framework Ever Authorise Real-Time Bidding After}} the
  {{Belgian DPA Decision}}?}
\newblock \bibinfo{journal}{\emph{Technology and Regulation}}
  \bibinfo{volume}{2022} (\bibinfo{year}{2022}), \bibinfo{pages}{12--22}.
\newblock
\urldef\tempurl%
\url{https://doi.org/10.26116/techreg.2022.002}
\showDOI{\tempurl}


\bibitem[Veale and Zuiderveen~Borgesius(2021)]%
        {vealeDemystifyingDraftEU2021}
\bibfield{author}{\bibinfo{person}{Michael Veale} {and}
  \bibinfo{person}{Frederik Zuiderveen~Borgesius}.}
  \bibinfo{year}{2021}\natexlab{}.
\newblock \showarticletitle{Demystifying the {{Draft EU Artificial Intelligence
  Act}}}.
\newblock \bibinfo{journal}{\emph{Computer Law Review International}}
  \bibinfo{volume}{22}, \bibinfo{number}{4} (\bibinfo{year}{2021}),
  \bibinfo{pages}{97--112}.
\newblock
\urldef\tempurl%
\url{https://doi.org/10/gns2s9}
\showDOI{\tempurl}


\bibitem[Verbeek(2005)]%
        {verbeek2005}
\bibfield{author}{\bibinfo{person}{Peter-Paul Verbeek}.}
  \bibinfo{year}{2005}\natexlab{}.
\newblock \bibinfo{booktitle}{\emph{{What Things Do: Philosophical Reflections
  on Technology, Agency, and Design}}}.
\newblock \bibinfo{publisher}{{Penn State Press}}.
\newblock


\bibitem[Wieringa(2020)]%
        {wieringa2020account}
\bibfield{author}{\bibinfo{person}{Maranke Wieringa}.}
  \bibinfo{year}{2020}\natexlab{}.
\newblock \showarticletitle{What to account for when accounting for algorithms:
  a systematic literature review on algorithmic accountability}. In
  \bibinfo{booktitle}{\emph{Proceedings of the 2020 conference on fairness,
  accountability, and transparency}}. \bibinfo{pages}{1--18}.
\newblock


\bibitem[Williams et~al\mbox{.}(2022)]%
        {williams2021}
\bibfield{author}{\bibinfo{person}{Rebecca Williams} {et~al\mbox{.}}}
  \bibinfo{year}{2022}\natexlab{}.
\newblock \showarticletitle{From transparency to accountability of intelligent
  systems: Moving beyond aspirations}.
\newblock \bibinfo{journal}{\emph{Data \& Policy}}  \bibinfo{volume}{4}
  (\bibinfo{year}{2022}).
\newblock


\end{thebibliography}
